\documentclass[12pt, draftclsnofoot, onecolumn]{IEEEtran}
\usepackage{multirow}
\usepackage[T1]{fontenc}
\usepackage{ifpdf}
\usepackage{cite}
\usepackage[pdftex]{graphicx}
\usepackage{amsmath}
\usepackage[cmintegrals]{newtxmath}
\usepackage{bm}
\usepackage{algpseudocode}
\usepackage{algorithm}
\usepackage{caption}
\usepackage{fixltx2e}
\usepackage{url}

\begin{document}

\title{A Combined Stochastic and Physical Framework for Modeling Indoor 5G\\ Millimeter Wave Propagation}

\author{\IEEEauthorblockN{Georges Nassif\IEEEauthorrefmark{1},
		Catherine Gloaguen\IEEEauthorrefmark{1}, and
		Philippe Martins\IEEEauthorrefmark{2}}\\ 
		\IEEEauthorblockA{\IEEEauthorrefmark{1}Orange Labs, France, \{georges.nassif, catherine.gloaguen\}@orange.com}\\
		\IEEEauthorblockA{\IEEEauthorrefmark{2}Telecom Paris, France, martins@telecom-paris.fr}}

\IEEEspecialpapernotice{\normalsize This work has been submitted to the IEEE for possible publication. Copyright may be transferred without notice, after which this version may no longer be accessible.}

\maketitle

\begin{abstract}
Indoor coverage is a major challenge for 5G millimeter waves (mmWaves). In this paper, we address this problem through a novel theoretical framework that combines stochastic indoor environment modeling with advanced physical propagation simulation. This approach is particularly adapted to investigate indoor-to-indoor 5G mmWave propagation. Its system implementation, so-called iGeoStat, generates parameterized typical environments that account for the indoor spatial variations, then simulates radio propagation based on the physical interaction between electromagnetic waves and material properties. This framework is not dedicated to a particular environment, material, frequency or use case and aims to statistically understand the influence of indoor environment parameters on mmWave propagation properties, especially coverage and path loss. Its implementation raises numerous computational challenges that we solve by formulating an adapted link budget and designing new memory optimization algorithms. The first simulation results for two major 5G applications are validated with measurement data and show the efficiency of iGeoStat to simulate multiple diffusion in realistic environments, within a reasonable amount of time and memory resources. Generated output maps confirm that diffusion has a critical impact on indoor mmWave propagation and that proper physical modeling is of the utmost importance to generate relevant propagation models.
\end{abstract}

\begin{IEEEkeywords}
5G, millimeter wave communication, indoor environment, stochastic geometry, physical propagation modeling, electromagnetic diffusion, link budget, memory management, optimization algorithms.
\end{IEEEkeywords}

\IEEEpeerreviewmaketitle

\section{Introduction}
Three use case families are defined for 5G mmWave applications. eMBB (enhanced Mobile Broadband) will deliver 10x higher data rates than 4G. mMTC (massive Machine Type Communications) will require high network capacity due to the huge number of connected devices. URLLC (Ultra-Reliable Low Latency Communications) will support applications with stringent latency and reliability constraints.
 
Although mmWaves are a promising solution to the spectrum scarcity problem, their propagation properties, namely their very short wavelength, can severely impact transmission. These impairments become even stronger in indoor-to-indoor communication scenarios due to numerous constraints imposed by the environment, leading to a new major challenge: indoor coverage.   

This is where indoor-specific studies on mmWave propagation modeling become essential for planning indoor 5G networks. This subject is, however, very complex to investigate since indoor environments vary a lot in terms of spatial configuration. Even when considering a particular geometry, its interior characteristics (furniture positions, doors, etc.) and materials (wood, concrete, etc.) are very diverse. Moreover, indoor material surfaces are generally rough; this roughness is comparable to the wavelength of mmWaves and thus, electromagnetic (EM) diffusion cannot be ignored and must be studied.

Existing investigations rely on measurement campaigns or system simulations. Conducting measurements is a heavy process and requires a lot of resources. The few studies \cite{Haneda, MacCartney} based on this method carry out measurements in a specific environment type and setting; collected data are accurate for that particular scenario only and do not account for the propagation diversity inherent to an environment's topology, morphology and material properties. Since it is impractical to conduct measurements in various environment configurations, generated empirical models cannot be used in other indoor mmWave applications, nor can they provide information regarding the influence of indoor characteristics on propagation mechanisms, especially diffusion.

System simulations offer a flexible alternative to predict propagation behavior. Existing indoor mmWave simulators allow users to study propagation in various environment settings. Each of these simulators uses different implementation techniques of propagation mechanisms; however, most of them suffer from major drawbacks when it comes to diffusion which is the most complex mechanism but certainly the one that should be mostly investigated. In \cite{remcom} for example, the diffusion lobe which physically describes the angular distribution of EM energy is oversimplified and set by the user, choosing between a Lambert or directive model, and whether or not to include backscattering, setting also the back lobe. Some simulators as \cite{winprop} utilize empirical models, and others as \cite{mmTrace} do not consider diffusion due to its implementation challenges. Hence, these simulators cannot predict the actual indoor mmWave behavior, nor can generated models provide conclusions regarding the impact of indoor geometry and materials on mmWave propagation.

In this paper, we present a novel modeling framework for 5G indoor mmWave propagation that combines stochastic indoor environment generation with advanced physical propagation simulation. Its system implementation, so-called iGeoStat, utilizes the theory of stochastic geometry to generate parameterized typical environments that account for the statistical indoor variability, and employs the complex but comprehensive physical model of He \cite{He} to simulate radio propagation based on the physical interaction between EM waves and indoor materials. Practical implementation of this framework raises highly challenging computational tasks that we solve by formulating an adapted link budget and designing new memory management and optimization algorithms. This makes iGeoStat the first to simulate multiple diffusion in realistic environments, allowing us to study the actual indoor behavior of 5G mmWave propagation.

The proposed framework aims to statistically understand the influence of the environment's geometry and materials' parameters on mmWave propagation properties, especially coverage and path loss. Therefore, generated models are not defined by the geometry or material itself, but rather by the parameters that characterize them. This framework is not dedicated to a particular environment, material, frequency or use case, making it very efficient for indoor 5G mmWave propagation modeling.

The contributions of this paper are summarized as follows:
\begin{enumerate}
	\item We introduce a new stochastic modeling approach that generates parameterized and realistic 3D environments, taking into account the indoor spatial variability.
	\item We present the first implementation of the physical model of He for radio propagation simulation based on the physical interaction between EM waves and indoor materials.
	\item We formulate an adapted link budget that enables multiple diffusion simulation and power tracking at any point in the environment.
	\item We design advanced algorithms for memory optimization, multiple diffusion management and simulation acceleration.
	\item We generate the first physical-based output maps (power, SINR, coverage, path loss and delay spread) for 5G Fixed Wireless Access at 60 GHz and industry 4.0 at 26 GHz.
\end{enumerate}

The rest of the paper is organized as follows. Section~\ref{sec:stochgeom} presents the theory behind the stochastic modeling of indoor environments. Section~\ref{sec:EM} presents the physical model of radio propagation simulation. The system implementation of our framework is presented in Section~\ref{sec:simulator}, along with the implementation challenges in Section~\ref{sec:impchal}. In Section~\ref{sec:results}, we evaluate the performance of iGeoStat for two major 5G applications, and validate our simulation results with existing measurement data. Finally, conclusions and perspectives are drawn in Section~\ref{sec:conclusion}.

\section{Stochastic Geometry}
\label{sec:stochgeom}
Stochastic geometry is a powerful tool used to generate mathematical models based on spatial probabilities \cite{Kendall}. It provides statistical information about the random configurations we wish to analyze. This tool is used in many disciplines like computer vision \cite{Nikolai}, wireless \cite{Haenggi} and fixed \cite{Catherine} networks, where analytic calculation and simulation can be done with few input parameters. 

Stochastic geometry incorporates the theory of random tessellations \cite{Chiu}, defined as the random division of space into convex non-overlapping polygonal regions; i.e. partitions. This section sketches the theory behind two random tessellations implemented in iGeoStat to generate typical parameterized indoor environments.

\subsection{Poisson Line Tessellation (PLT)}
\label{subsec:PLT}
A PLT is a set of random lines driven by an underlying Poisson Point Process, where each point represents the origin of a line drawn according to an angular distribution in $[0,\pi[$. The associated PLT in the plane is a set of random lines delimiting polygonal cells (Fig.~\ref{fig:PLTSTIT}), characterized by its intensity $L_A$, i.e. the mean total edge length per unit area. Anisotropy can be described by a probability distribution $\mathcal{R}$ in the space of directions and is measured via the anisotropy parameter $\xi$ (Eq.~\ref{eq:aniso}), where $|\sin\measuredangle(u,v)|$ is the area of a parallelogram drawn by unit vectors $u$ and $v$ \cite{Nagel1}:
\begin{equation}\label{eq:aniso}
\xi=\iint|\sin\measuredangle(u,v)|\mathcal{R}(du)\mathcal{R}(dv)
\end{equation}
The isotropic case corresponds to uniform $\mathcal{R}$ and $\xi=2/\pi$. The simulator uses an anisotropy coefficient $\alpha$ derived from $\xi$ and defined between 0 (isotropic) and 1 (anisotropic).

\subsection{STable by ITeration tessellation (STIT)}
\label{subsec:STIT}
\begin{figure}[h]
	\centering
	\includegraphics[width=16.5 cm]{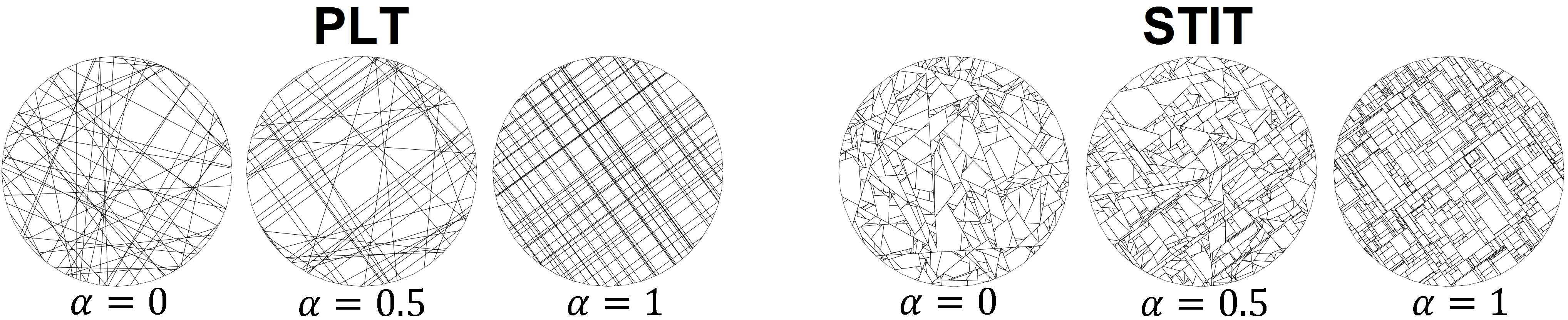}
	\caption{Realizations of PLT and STIT tessellations for various anisotropy coefficients.}\label{fig:PLTSTIT}
\end{figure}

The STIT tessellation introduced in \cite{Nagel} is indexed with time. The initial window has a random exponentially distributed lifetime $t$ after which it is divided into two new cells by a random line. These new cells are each attributed independently a lifetime that drives their division by a new random line. The parameter of the time probability distribution is chosen here inversely proportional to the cell perimeter so that larger cells tend to die sooner. The whole process is stopped at an arbitrary time that can be transformed into the mean total edge length $L_A$ depending on the anisotropy. The result (Fig.~\ref{fig:PLTSTIT}) is a STIT formed of polygons that do not depend on the initial window and is parameterized by $L_A$. Anisotropy $\xi$ can be also considered here. We note that the interior of a typical STIT ($L_A,\xi$) cell has the same distribution as a PLT ($L_A,\xi$) one.

\subsection{Morphological features}
The morphology of a stationary planar tessellation is described by the mean values of the number of nodes, cells, etc. per unit area. Mean value formulae \cite{Nagel1} exist for PLT and STIT topologies (Table~\ref{tab:tab2}), allowing us to fit the tessellation parameters to real data as in \cite{GloaguenFit}.
\begin{table}[h]
	\caption{Mean values for PLT and STIT tessellations ($L_A,\xi$).}
	\label{tab:tab2}
	\centering
	\begin{tabular}{|c|c|c|}
		\hline 
		\textbf{Morphology} & \textbf{PLT} & \textbf{STIT} \\ 
		\hline
		Total edge length & $L_A$ & $L_A$ \\
		\hline
		Number of vertices & $\frac{1}{2}\xi L_A^2$ & $L_A^2\xi$ \\
		\hline
		Number of edges & $L_A^2\xi$ & $\frac{3}{2}L_A^2\xi$ \\
		\hline
		Number of cells & \multicolumn{2}{c|}{$\frac{1}{2}L_A^2\xi$} \\
		\hline
		Length of a typical edge & \multicolumn{2}{c|}{$2/(3L_A\xi)$} \\
		\hline
		Perimeter of a typical cell $\bar{P}$  & \multicolumn{2}{c|}{$4/(L_A\xi)$} \\
		\hline
		Area of a typical cell $\bar{A}$ & \multicolumn{2}{c|}{$2/(L_A^2\xi)$} \\
		\hline		 
	\end{tabular}
\end{table}

\section{EM Propagation Modeling}
\label{sec:EM}
This section presents the propagation model implemented in iGeoStat to simulate reflection and diffusion. Reflection occurs when a wave is incident on a smooth surface; the outgoing `specular' direction is given by Snell-Descartes law and the fraction of reflected power is provided by Fresnel's coefficients. Diffusion, on the other hand, is more complex and occurs when a wave is incident on a rough surface. This mechanism is crucial for indoor mmWave propagation and must not be neglected. Its behavior should not be user-defined but solely based on the physical interaction between EM waves and material properties.

\subsection{Diffusion from rough surfaces}
\begin{figure}[h]
	\centering
	\includegraphics[width=7.5 cm]{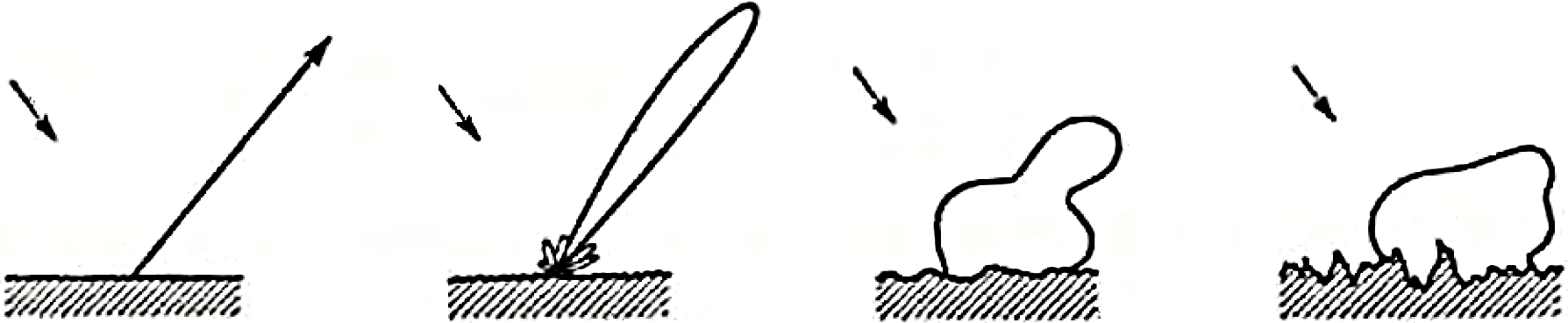}
	\caption{Impact of increasing surface roughness (left to right) on the diffusion lobe (from \cite{Beckmann}).}
	\label{fig:reftodiff}
\end{figure}

Surface roughness is defined here with respect to the incident wavelength. Typical indoor materials (carpets, furniture, etc.) have rough surfaces with regards to mmWaves. Fig.~\ref{fig:reftodiff} illustrates the transition from specular reflection to diffusion, induced by increasing surface roughness.

The problem of diffusion from rough surfaces is addressed according to three different approaches. (I) Empirical ones reproduce measured data without any physical interpretation \cite{Blinn,Lambert,Phong,Ward}. (II) Geometrical ones use ray optics assuming that irregularities are much larger than the wavelength \cite{Torrance,Oren,Ashikmin}. (III) Physical ones apply EM theory; they are very complex but more general \cite{Beckmann,He}. We note that physical diffusion models are used in computer graphics and video games \cite{Berg}, but have never been implemented to simulate radio propagation.

In the context of (III), we chose to implement the physical model of He \cite{He} since it encompasses all the other models, supports different types of surface roughness and includes all major EM propagation mechanisms: reflection, diffusion, surface diffraction, interference, polarization, masking and shadowing. Although developed for visible light, its physical approach allows it to be used for any frequency.

\begin{figure}[h]
	\centering
	\includegraphics[width=7.5 cm]{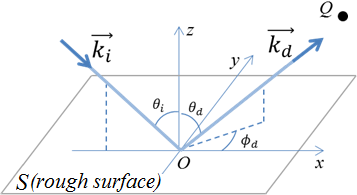}  
	\caption{Geometry of the diffusion problem.}
	\label{fig:geometry}
\end{figure}

The geometry of the diffusion problem is illustrated in (Fig.~\ref{fig:geometry}). An EM wave propagating at a wavelength $\lambda$ is incident on a rough interface $S$ separating the air and a medium of complex refractive index $n_c(\lambda)$. $(0,x,y,z)$ are the natural Cartesian coordinates defined by the surface's average plane ($z=0$) and the incident plane wave vector $k_i$. We are interested in finding the diffused intensity at any observation point $Q$ in the far field ($\theta_d\in~[0,\frac{\pi}{2}]$; $\phi_d\in[0,2\pi]$) where the diffused wave can be assumed planar with wave vector $k_d$. The illuminated area is $L_x\times L_y$ such that $L_x,L_y \gg \lambda$. The polarization directions $\vec{s}$ and $\vec{p}$, associated to a planar wave $\vec{k}$ are respectively in the $z=0$ and $(\vec{k},\vec{z})$ planes.

\subsection{He's Model}
\label{subsec:he}
\begin{figure}[h]
	\centering
	\includegraphics[width=8 cm]{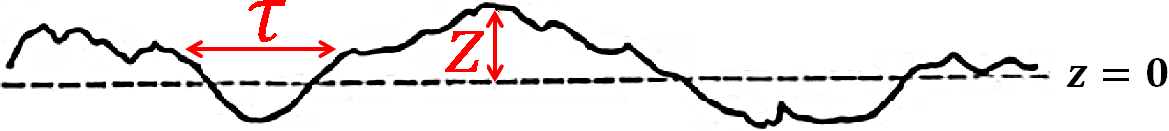}
	\caption{Random surface profile.}\label{fig:surfprof}
\end{figure}

He's model solves the diffusion problem using the Kirchhoff Approximation integral (KA) \cite{Beckmann} (which only holds for very large surface curvatures compared to the wavelength), coupled with a random process modeling of the surface irregularities, since they cannot be accurately described. The model assumes that the height distribution $z=\zeta(x,y)$ is spatially isotropic and Gaussian of zero mean value and $\sigma_0$ standard deviation. The dimensionless parameter $\sigma_0/\lambda$ defines the surface roughness. Height values taken at two points on the surface are separated by an autocorrelation distance $\tau$ (Fig.~\ref{fig:surfprof}).  Using this method, He's model provides the statistical average of the diffused intensity $I^d(Q)$ over the realizations of the random surface $S$, through the Bidirectional Reflectance Distribution Function (BRDF) \cite{Ticconi,Ulaby}.

\subsection{BRDF}
\label{subsec:rho}
The BRDF, denoted $\rho$, represents the spatial distribution of $I^d(Q)$ and is defined as the ratio of total diffused intensity in the direction $(\theta_d,\phi_d)$ to incident intensity on the surface element in the direction $\theta_i$.

\begin{figure}[h]
	\centering
	\includegraphics[width=6.5 cm]{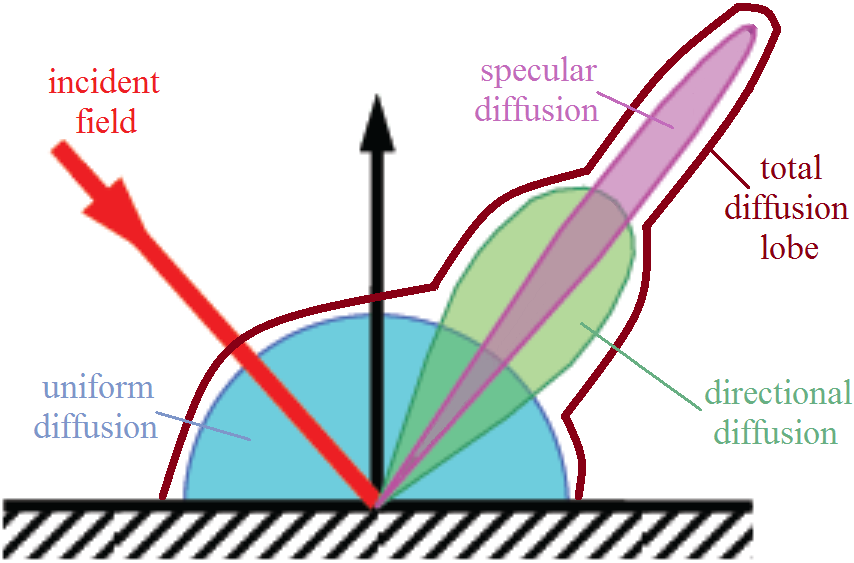}
	\caption{The BRDF components.}\label{fig:brdf}
\end{figure}

$\rho$ is the sum of $\rho_{sr}$ (specular), $\rho_{dd}$ (directional diffuse) and $\rho_{ud}$ (uniform diffuse) (Fig.~\ref{fig:brdf}):
\begin{equation}
\rho=\frac{dI^d(\theta_d,\phi_d)}{I^i(\theta_i)\cos\theta_id\omega_i}=\rho_{sr}+\rho_{dd}+\rho_{ud}
\end{equation}
The first two components result from first-order surface reflections on the tangential planes: $\rho_{sr}$ is due to reflection from the mean surface and $\rho_{dd}$ is due to diffraction scattering from surface irregularities. The third component $\rho_{ud}$ results from multiple surface and subsurface reflections. We note that due to transmission losses in the lower half-space, the integration of $\rho$ over the upper-half space should verify $\iint_{[2 \pi]} \rho \sin\theta_d d\theta_d d\phi_d \leq1$.

He's model provides the $\rho$ expression for a polarized incident fields ($pol=\vec{s}$ or $\vec{p}$):
\begin{equation}\label{eq:rhopol}
\rho_{pol}=\frac{\mathcal{S}(\mathcal{N}_1+\mathcal{N}_2)(\mathcal{F}_s+\mathcal{F}_p)}{(4\pi)^2\cos\theta_i\cos\theta_d}=\rho_{sr}+\rho_{dd}
\end{equation}
$\mathcal{S}$ denotes a shadowing and masking function, $\mathcal{N}_1$ is the specular contribution due to reflection from the mean surface, and $\mathcal{N}_2$ is the directional diffuse one due to diffraction scattering from surface irregularities. $\mathcal{F}_{s}$ and $\mathcal{F}_{p}$ describe cross-polarization effects due to the orientation of the $\zeta$ height field and are functions of the Fresnel coefficients for the angle bisector of $\vec{k}_i$ and $\vec{k}_d$. The missing third component $\rho_{ud}$ is generally approximated by a function $a(\lambda)$, estimated from experiments \cite{Katsushi}. In our case, it is deduced approximately from an estimation of the transmitted component that ensures energy conservation (Section~\ref{subsec:brdfcomp}).

For non-specular directions the field is always incoherent \cite{Beckmann} and it is impossible to reconstruct the polarization vector of the diffused field. Hence the unique way to consider a series of multiple diffusion is to assume non-polarized incident field and use power link budget (Section~\ref{subsec:radpropag}). The non-polarized BRDF is defined as the statistical average of $\rho_{pol}$ considering an incident polarization angle uniformly distributed in $[0, 2 \pi]$. Noting $\mathcal{F}_{nonp}=<\mathcal{F}_s+\mathcal{F}_p>$:
\begin{equation}\label{eq:rho}
\rho_{nonp}=<\rho_{pol}>=\frac{\mathcal{S}(\mathcal{N}_1+\mathcal{N}_2) \mathcal{F}_{nonp}}{(4\pi)^2\cos\theta_i\cos\theta_d}
\end{equation}
Detailed analytic expressions of  $\mathcal{S}, \mathcal{N}_1, \mathcal{N}_2$, $\mathcal{F}_{s,p}$ and $\mathcal{F}_{nonp}$ are found in \cite{He}. Their numerical computation is far from being straightforward and was performed as a first step in Mathematica \cite{Mathematica} for validation purposes by comparing the results to figures from papers \cite{He,Tariq,Bass,Beckmann}.

The input parameters to compute $\rho$ are presented in Table~\ref{tab:brdfparam}. The wavelength $\lambda$ defines the reference length scale. The material's complex refractive index $n_c(\lambda)$ required for the Fresnel coefficient is not usually known a priori; measurement values can be found in \cite{Sato, Li} for some typical indoor materials, measured at mmWave bands.

Fig.~\ref{fig:visubrdf} illustrates spherical 3D plots of $\rho_{nonp}$ that correspond to parameters (Table~\ref{tab:brdfparam}) inspired by He's example of roughened aluminum. These plots allow us to visualize how the incident energy diffuses in space as a function of ($\theta_d,\phi_d$). The effect of an increase in surface roughness is illustrated by the transition from a near specular reflection to a large diffusion lobe. As roughness increases, the contribution of $\rho_{sr}$ diminishes with respect to $\rho_{ud}$. The case of a smooth surface ($\sigma_0=0$) is naturally included in $\rho_{nonp}$ and does not require specific treatment.
\begin{table}[h]
	\begin{minipage}[c]{0.45\linewidth}
		\caption{Input parameters for (a) slightly rough, (b) moderately rough and (c) very rough surfaces. $\theta_i=30^{\circ}$, $\theta_d\in[0,\frac{\pi}{2}]$ and $\phi_d\in[0,2\pi]$}
		\label{tab:brdfparam}
		\centering
		\begin{tabular}{|c|c|c|c|}
			\hline 
			\textbf{Parameters} & \textbf{(a)} & \textbf{(b)} & \textbf{(c)} \\ 
			\hline
			$\lambda\,(\mu m)$ & 2 & 1 & 2 \\
			\hline
			$\sigma_0\,(\mu m)$ & 0.3 & 0.3 & 1.5 \\
			\hline
			$\sigma_0/\lambda$ & 0.15 & 0.3 & 0.75 \\
			\hline
			$\tau\,(\mu m)$ & 2 & 2 & 1.8 \\
			\hline
			$n_c$ & $3+21i$ & $3+10i$ & $3+21i$ \\
			\hline
		\end{tabular}
	\end{minipage}\hfill
\begin{minipage}[c]{0.5\linewidth}
	\centering
	\includegraphics[width=8.2 cm]{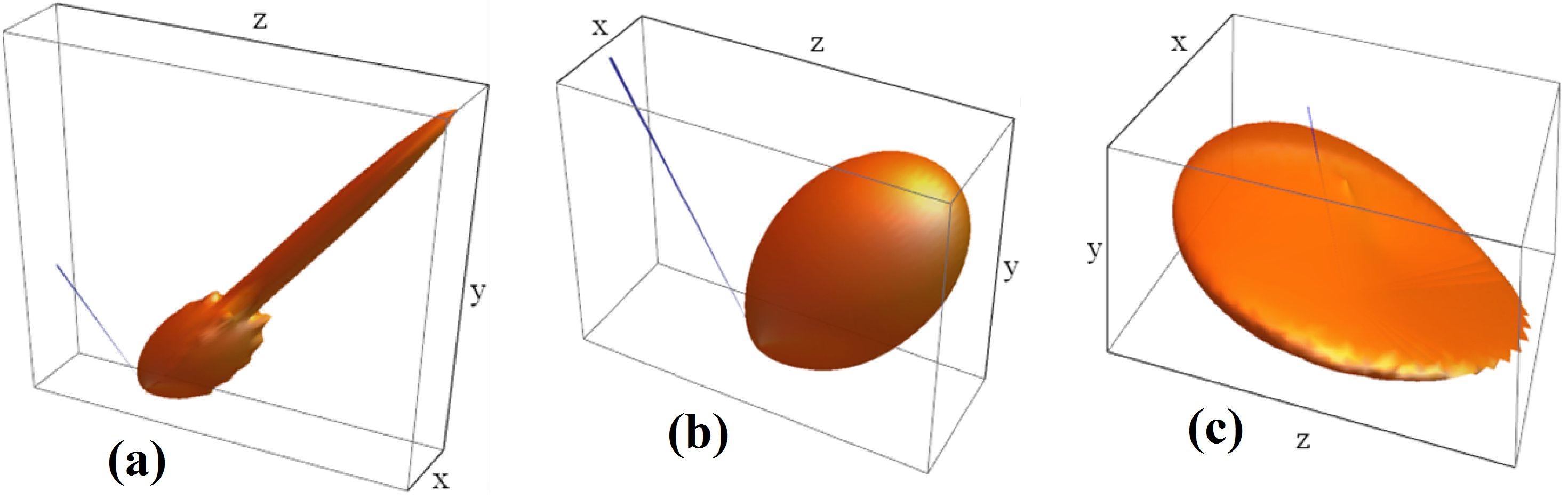}
	\captionof{figure}{3D plots of $\rho_{nonp}$ computed with Mathematica for input parameters in (Table~\ref{tab:brdfparam}). The incoming direction is materialized by the thin blue pencil.}
	\label{fig:visubrdf}
\end{minipage}
\end{table}

We note that the incidence angle $\theta_i$ has a strong impact on the directivity of $\rho_{nonp}$, tending to specularity as $\theta_i$ increases. For the sake of simplicity, $\rho_{nonp} + a(\lambda)$ is hereafter denoted by $\rho$.

\section{System Implementation} \label{sec:simulator}
The system implementation of our framework, so-called iGeoStat, is written in C++ and consists of three main modules: the first one (Section~\ref{subsec:indenv}) generates parameterized indoor environments based on the random tessellations discussed in Section~\ref{sec:stochgeom}, the second one (Section~\ref{subsec:radpropag}) simulates radio propagation according to the mechanisms presented in Section~\ref{sec:EM}. From this, the third one generates various output maps based on the measurement plane (Section~\ref{subsec:output}). Note that all parameters below are user-defined in the iGeoStat configuration file. 

\subsection{Generating parameterized indoor environments}
\label{subsec:indenv}
This module uses stochastic geometry and random tessellations (Section~\ref{sec:stochgeom}) to generate various typical 3D indoor environments like apartments, commercial centers, open-space offices, etc., using a minimum number of input parameters.

\textbf{Generating a random tessellation:} the simulator starts by defining the boundaries of the area on which to build the environment. The reference system is the orthogonal 2D Cartesian coordinate system (Fig.~\ref{fig:gentess}) plus a vertical coordinate $h$. The `initial cell' represents the common floor, centered at the origin. Its polygonal shape is defined by the radius of its circumscribed circle and the number of its sides. This cell is then divided into sub-cells according to a PLT or STIT topology, choosing as reference the area $\bar{A}$ or perimeter $\bar{P}$ of a typical cell (Table~\ref{tab:tab2}). To generate realistic floor plans with rectangular partitions, we set the anisotropy coefficient to $\alpha=1$. The tessellation can be then rotated by a tilt angle with respect to the x-axis (Fig.~\ref{fig:gentess}(c)).
\begin{table}[h]
	\begin{minipage}[c]{0.45\linewidth}
		\caption{Input parameters to generate the tessellations in Fig.~\ref{fig:gentess} (arbitrary units).}
		\label{tab:tessparam}
		\centering
		\begin{tabular}{|c|c|c|c|}
			\hline 
			\textbf{Parameters} & \textbf{(a)} & \textbf{(b)} & \textbf{(c)} \\ 
			\hline
			Number of sides & 4 & 5 & 6 \\
			\hline
			Window radius & 300 & 300 & 300 \\
			\hline
			Topology & PLT & STIT & STIT \\ 
			\hline 
			Morphology & area & perimeter & area \\ 
			\hline
			Mean value & 1700 & 50 & 1000 \\
			\hline
			Tilt angle & 0 & 0 & $35^\circ$ \\
			\hline
		\end{tabular}
	\end{minipage}\hfill
	\begin{minipage}[c]{0.5\linewidth}
		\centering
		\includegraphics[width=8.2 cm]{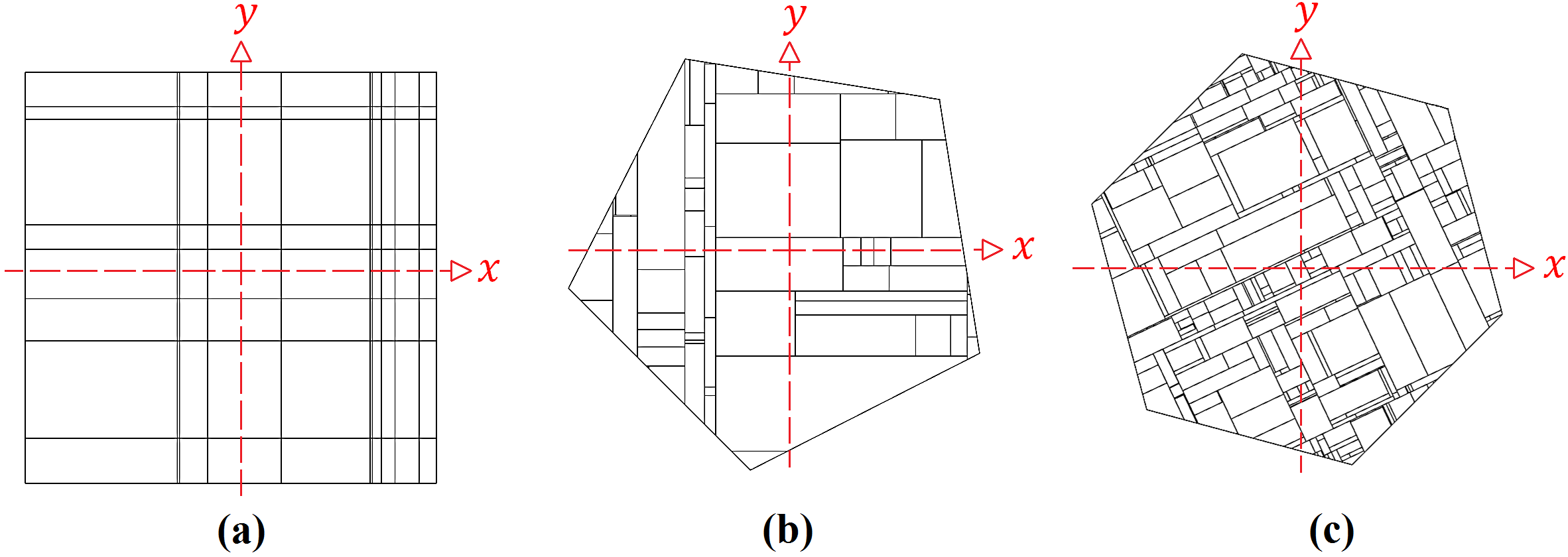}
		\captionof{figure}{Generated tessellations based on parameters in Table~\ref{tab:tessparam}.}
		\label{fig:gentess}
	\end{minipage}
\end{table}

All the cells, including the initial one, are then stored in a vector of cells as a circular linked list of edges. Each edge is a 2D coordinates vector of its `start' and `end' vertices, where the `start' vertex of the first edge is the `end' vertex of the last one.

\textbf{Drawing a 2D floor plan:} the cells have to be modified to represent the layout of the different rooms or sub-spaces in the indoor environment, separated by corridors. The simulator proceeds by iterating through the vector of cells, creating inside each one of them a polygon of equal number of parallel sides, distant by a half corridor width $w_c$ (minus sampling technique \cite{Chiu}). Segment lines delimiting the cells are then removed to visualize the spacing between adjacent polygons, now separated by the width of a corridor.

\begin{figure}[h]
	\centering
	\includegraphics[width=13 cm]{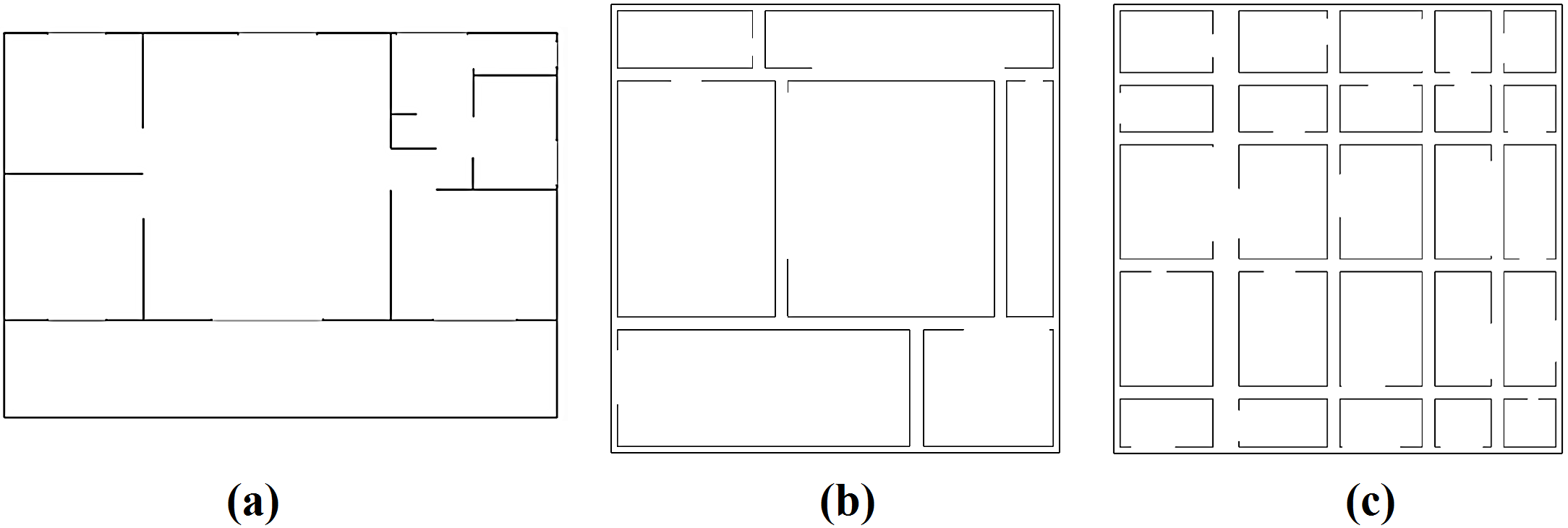}
	\caption{(a) Real layout of an indoor environment. (b) and (c) 2D floor maps generated from random STIT and PLT tessellations.}
	\label{fig:floorplan}
\end{figure}

In order to represent doors or entrances of the indoor rooms, a single opening is added to a randomly selected edge of center $m$ and length $l$. The user chooses among two options `pcent' or `rand'. With `pcent' and parameter $w_0$ (\%), a door of width $w_0 l$ centered at $m$ is drawn. With `rand', a door is drawn with a uniformly random $w_0 l$ around $m$. These modifications transform a random tessellation from an arrangement of polygons to a typical 2D floor plan (Fig.~\ref{fig:floorplan}).

\textbf{Transforming a 2D floor plan into a 3D indoor environment:} the whole indoor environment (initial cell) is closed by adding surrounding walls and a common ceiling at height $h_c$ and thus becomes a `vertical polygon'. The same thing is done for each of the polygons delimiting indoor rooms or sub-spaces, the inner walls height is defined by $h_{wa} < h_c$. Since the rooms are separated from their neighbors by a corridor it is possible to assign a random height to each of them, to mimic open spaces for example. In this case, the height is drawn from an exponential distribution of the form $h_{wm}+\exp[1/(h_{wa}-h_{wm})]$ of average $h_{wa}$, minimum $h_{wm}$ and truncated to $h_c$.

Fig.~\ref{fig:3Dindenv} illustrates parameterized 3D environments generated based on Table~\ref{tab:envparam}, highlighting the influence of varying structuring parameters on the indoor geometry. Interpreting an environment's structure depends on the user's perspective; nevertheless, this shows our ability to generate parameterized typical indoor environments based on stochastic geometry.

Generating a parameterized 3D indoor environment is a seamless single-step process that takes up to 2 milliseconds and 10 KB in memory usage on a low-end hardware configuration (a 2.2 GHz Intel Core i3-350M CPU with 4 GB of RAM).
\begin{table}[h]
	\begin{minipage}[c]{0.45\linewidth}
		\caption{Input parameters to generate the various 3D indoor environment of Fig.~\ref{fig:3Dindenv} (arbitrary units)}
		\label{tab:envparam}
		\centering
	\begin{tabular}{|c|c|c|c|c|}
		\hline 
		\textbf{Parameters} & \textbf{(a)} & \textbf{(b)} & \textbf{(c)} & \textbf{(d)} \\ 
		\hline 
		Topology & STIT & PLT & STIT & STIT \\ 
		\hline 
		$h_c$ & 10 & 10 & 10 & 10 \\
		\hline
		$w_c$ & 25 & 30 & 35 & 40 \\ 
		\hline 
		$w_o$ & rand & rand & 40\% & rand \\
		\hline
		$h_{wa}$ & 5 & 10 & 10 & 2 \\ 
		\hline
		$h_{wm}$ & - & - & 5 & - \\
		\hline 
	\end{tabular}
	\end{minipage}\hfill
	\begin{minipage}[c]{0.5\linewidth}
		\centering
		\includegraphics[width=8.2 cm]{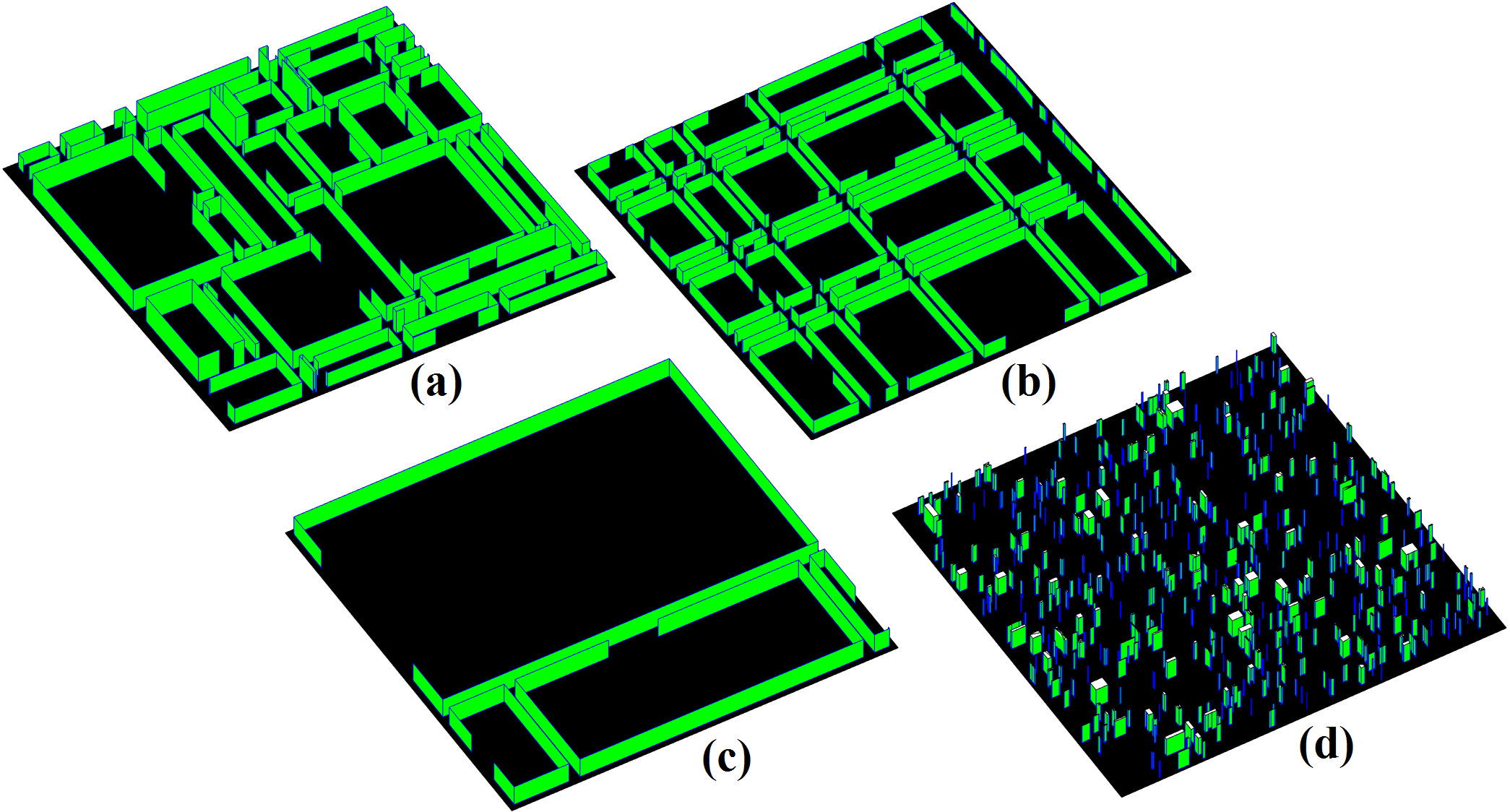}
		\captionof{figure}{3D environments that may represent: (a) open-space office, (b) commercial center, (c) sports center, (d) book fair (people and stands).}
		\label{fig:3Dindenv}
	\end{minipage}
\end{table}

A custom mode for generating indoor environments is also implemented in the simulator. In contrary to the random mode, it allows users to generate a personalized environment based on given 3D coordinates data. The above mentioned fitting methods \cite{GloaguenFit} can determine the tessellation parameters that would best fit a given deterministic environment.

\subsection{Radio propagation simulation}
\label{subsec:radpropag}
This module simulates radio propagation in the generated 3D indoor environment, where diffusion is simulated according to He's physical model (Section~\ref{subsec:he}).

\textbf{Propagation methodology:} to consider diffusion, it is fundamental to keep track of the $(\theta_d, \phi_d)$ angular distribution of $\rho$ (Section~\ref{subsec:rho}) at each impact point, which is strongly dependent on $\theta_i$. Our implementation is based on the discretization of diffusion directions, with steps $\Delta\theta_d$ and $\Delta\phi_d$. Each diffusion direction ($\theta_d, \phi_d$) is attributed a solid angle and diffused power, enabling straightforward power tracking through an adapted link budget. Diffusion is thus monitored at each impact in a physical way; this allows us to reduce the set of ($\theta_d, \phi_d$) directions by adapting discretization steps or adding thresholds on $\rho$ and power values (Section~\ref{subsec:diffmanage}).

\textbf{Antenna parameters:} the antenna is placed in the reference 3D Cartesian coordinate system of the environment at $(x_a, y_a, h_a)$. Its position can be user-defined with respect to the geometry of a deterministic environment, or randomly set in a stochastic one. This approach aims to derive statistics on various propagation indicators related to the global parameters of the environment.

By assumption, the antenna emits uniformly a total power $P_0$ at wavelength $\lambda$ in a solid angle $\Omega_a$ centered in the spherical direction  ($\theta_0^a$ in azimuth, $\phi_0^a$ in elevation). Using dedicated discretization steps $\Delta \theta_0$ and $\Delta \phi_0$, $\Omega_a$ is divided into $N_0$ elementary solid angles $d^2 \Omega_0$, each carrying in its own direction $(\theta_0, \phi_0)$ a power $P_1=\frac{P_0 d^2 \Omega_0}{\Omega_a}$ to be included in the link budget. 

\begin{figure}[h]
	\centering
	\includegraphics[width=8 cm] {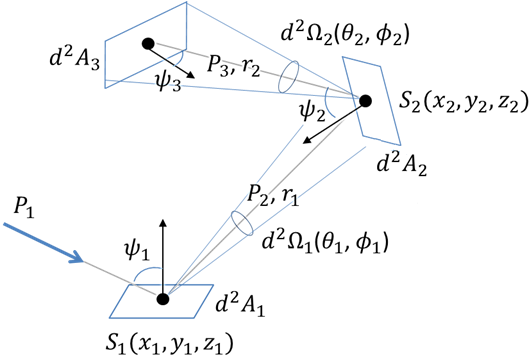}
	\caption{Power tracking after multiple diffusion.}\label{fig:pmultiple}
\end{figure}

\textbf{BRDF:} the BRDF from He's model is implemented separately from the main core of iGeoStat using the mathematical library PARI \cite{parigp}. To avoid computing $\rho$ for the same input parameters over and over again, numerical simulations of $\rho$ are executed for a whole set of input parameters and stored in databases using SQLite \cite{sqlite}. Hence, at each impact point, the incoming direction and surface normal define the incidence angle $\psi_j$ (Fig.~\ref{fig:pmultiple}) and the $\rho$ value of each diffusion direction ($\theta_d, \phi_d$) is retrieved from the databases.

Each BRDF database corresponds to computing $\rho$ for a single material ($\sigma_0$, $\tau$, $n_c$), wavelength ($\lambda$), and all discretized incident angles ($\theta_i$), and diffusion directions ($\theta_d$, $\phi_d$). Further details concerning the challenges related to He's model implementation, and the optimization of $\rho$ computation and storage are discussed in Section~\ref{subsec:brdfcomp}.

\textbf{Link budget:} using $\rho$, we can keep track of the power at any point of the space after multiple diffusion on various rough surfaces. In a global reference coordinate system $(0,x,y,z)$, let us consider a `diffusion point', or `impact point', i.e. an infinitesimal surface element $d^2 A_1$ (m$^2$) centered at $S_1 (x_1,y_1,z_1)$. $P_1$ (W) denotes the power received by $d^2 A_1 $ from some source in the far field direction $\psi_1$ with respect to its normal (Fig.~\ref{fig:pmultiple}).

The power diffused in the solid angle $d^2 \Omega_1$ in the direction $(\theta_1,\phi_1)$ is $P_2=P_1 \rho_1 d^2 \Omega_1$~(W), where $\rho_1(\psi_1, \theta_1,\phi_1)$ is the non-polarized BRDF computed for the surface parameters, the corresponding angles, and wavelength $\lambda$. The direction $(\theta_1,\phi_1)$ from $S_1$ defines a point $S_2(x_2,y_2,z_2)$ (assumed to exist) at distance $r_1$ on another planar surface, as well as angle $\psi_2$ with respect to the normal at $S_2$. The relationship between the illuminated surface $d^2 A_2$ and the solid angle $d^2 \Omega_1$ allows us to write $P_2$ as a function of surface element and distance as $P_2= P_1 \rho_1  d^2 A_2 \cos\psi_2 /r_1^2$. The power received on an infinitesimal surface after $j$ diffusion is then given by:
\begin{equation}\label{eq:power}
P_j=P_1 \prod_{i=1}^{j} \rho_j(\psi_j, \theta_j,\phi_j)  \frac{d^2 A_{j+1}  \cos\psi_{j+1} }{r_j^2}  
\end{equation}
Gas attenuation is not relevant here, but transmission and absorption losses in material are taken into account by $\rho$.

\textbf{Trajectory tracking:} a customized hierarchical data structure (Section~\ref{subsec:trackdiff}) is implemented to track `trajectories', i.e. the history of all impacts, diffusion directions, etc. To minimize the time required to compute diffusion directions, the environment is first recursively divided, a parameterized number of times into four sub-regions called quadrants. Every quadrant is a container of the different obstacles that exist in that region, i.e. an `obstacle vector' of floor, ceiling, walls, and measure (measurement plane), along with their geometrical information. The propagation then kicks off at the antenna from a set of $N_0\;d^2\Omega_0$. Each solid angle $d^2 \Omega_0$ intercepts and illuminates a small area on an indoor obstacle. Depending on the quadrant where the impact point occurs, the obstacles vector is iterated to deduce its corresponding  type and material.

\begin{figure}[h]
	\centering
	\includegraphics[width=7 cm]{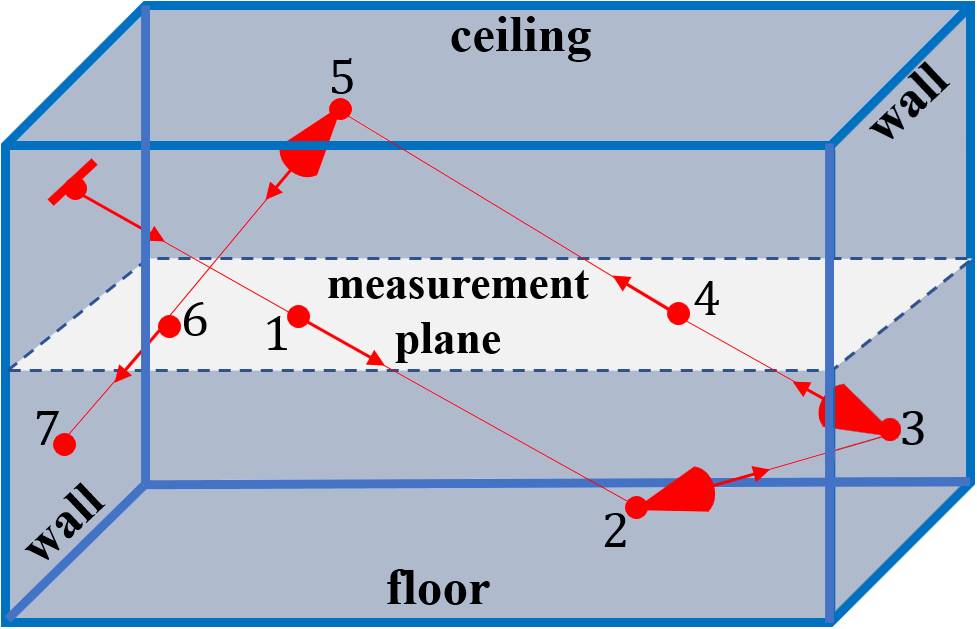}
	\caption{Trajectory representation of a unique direction from the antenna hitting 7 obstacles. (2-3-5-7) are impacts and (1-4-6) are measurement points.}
	\label{fig:measplane}
\end{figure}

Fig.~\ref{fig:measplane} sketches an example of a unique trajectory. Usually, a great number of diffusion directions are generated at each impact point, thus the necessity of an adapted data structure to ensure accurate tracking and storage of all propagation related information, and ultimately the possibility to generate various output maps. Each trajectory is a vector of pointers; each pointer corresponds to an impact point and contains the following information: impact's position and index, obstacle's type, total traveled distance of the trajectory, incidence angle and vector of outgoing diffusion directions with their corresponding power, solid angle and $\rho$.

\subsection{Measurement plane and simulation output}
\label{subsec:output}
This module generates various output maps as simulation outputs based on a user-defined measurement plane.

\textbf{Measurement plane:} it is imaginary, horizontal, parallel to the floor at height $h_m$ and covers the whole indoor environment. This plane aims to identify all the diffusion directions (and their propagation information) that cross it in order to generate the output maps. The intersection points are called `measurement points' to differentiate them from impact points where diffusion occurs. Since this plane is an imaginary obstacle, propagation directions crossing it are unaltered.

Measurement points are also pointers that contain the same information as a regular impact point, only that these information are evaluated and stored at the measurement plane and used to generate various output maps such as the impacts map, received power, Signal to Interference-plus-Noise Ratio (SINR), antenna coverage and delay spread.

\textbf{Simulation output:} the propagation simulation stops when the user-defined maximum number of multiple diffusion is reached for each of the $N_0$ antenna trajectories. To generate the output maps, the measurement plane is covered by a squared or circular grid of $n_1\times n_2$ resolution. This fixes the number of rectangular elements in length and width, or the radial and angular arrangement of sector areas. Grid elements are then filled with their corresponding measurement points based on their positions over the measurement plane and output maps are generated by extracting the relevant information from all the points in each grid element.

The grid resolution must be adapted based on a compromise between precision and visualization. A low resolution gives a continuous map over the environment with smooth variations in values or corresponding colors, but the output is not very precise as each grid element covers a large environment area. A high resolution may illustrate abrupt changes in the map values or colors, but the output is very precise as each grid element covers a small environment area.

\section{Implementation Challenges}
\label{sec:impchal}
We present in this section the main challenges of implementing the data structure that manages diffusion, as well as the challenges of computing, storing and retrieving BRDF values and preventing the explosion of diffusion directions. 

\subsection{Tracking impact points}
\label{subsec:trackdiff}
As discussed in Section~\ref{subsec:radpropag}, when diffusion occurs, multiple propagation directions are launched after every impact. Each direction corresponds to a potential source of diffusion or new impact point. The main challenge is to implement a data structure that keeps track of the impact points in order to accurately compute their corresponding propagation information (received power, traveled distance, etc.) when undergoing multiple diffusion.

\textbf{Choosing the optimal data structure:} several structures were investigated: the linear ones (arrays, linked lists, etc.) and the hierarchical ones (trees, graphs, etc.). However, the dependency of an impact point on its previous one induces a parent $\longrightarrow$ child relationship, which mostly correspond to a tree structure.
\begin{figure}[h]
	\centering
	\includegraphics[width=7.5 cm]{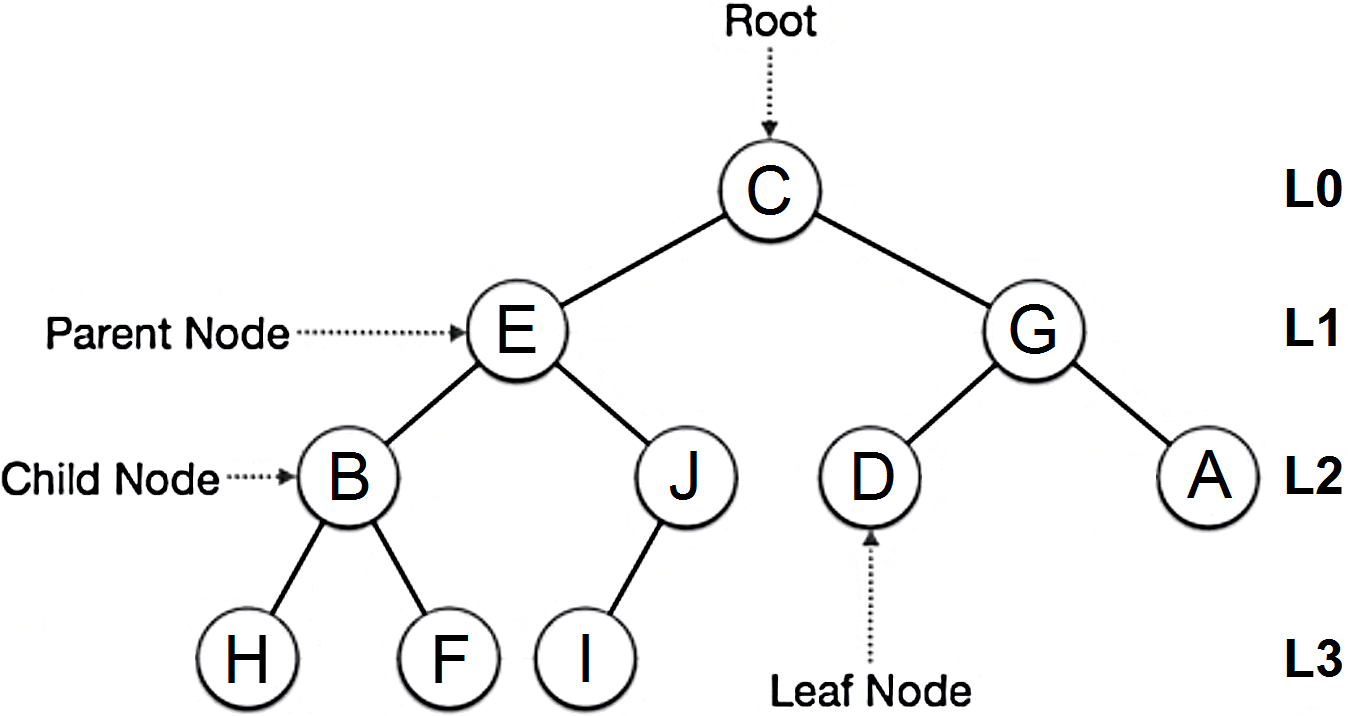}
	\caption{Main components of a tree data structure.}
	\label{fig:tree}
\end{figure}

The analogy between the tree components (Fig.~\ref{fig:tree}) and diffusion is as follows: the root and its edges represent the antenna and its trajectories. A parent node and its outward edges correspond to an impact point and its outgoing propagation directions. Each child node represents the resulting impact point from these directions and a leaf node corresponds to the last impact point of an antenna's trajectory. At each diffusion, a child node becomes a new parent node.

\textbf{Traversing a tree structure:} The challenge that arises when implementing a tree is the choice of its traversal, i.e. the order in which the tree nodes are treated. In our case, treating a node means computing power, distance, diffusion directions, etc. at that impact point. We recall that an impact point is a pointer containing all the propagation information. Several tree traversals were investigated (Depth-first (DFS), Breadth-First (BFS), Monte Carlo (MCTS), etc.).

BFS is a corecursive level-order algorithm, i.e. it uses self-produced data bit-by-bit as they become available to produce further bits of data. It is implemented using a queue; starting from the root, tree nodes are explored by visiting horizontally all the current level adjacent nodes before moving to the next level. Referring to Fig.~\ref{fig:tree}, this is equivalent to the following visiting order: C-E-G-B-J-D-A-H-F-I. BFS has a time complexity equal to $O(|N|+|E|)$, and a space complexity equal to $O(|N|)$, where $N$ is the number of nodes and $E$ the number of edges.

A customized BFS (Algorithm~\ref{alg:bfsalg}) was implemented as it corresponds the most to the logic of multiple diffusion. Starting from the antenna, all the adjacent parent impacts must be treated before any of the child impacts. Algorithm~\ref{alg:bfsalg} integrates a `Garbage Collection' (GC) that treats antenna trajectories independently from each other, allowing each one of them to reuse the same memory resources regardless of the number of antennas and trajectories $N_0$.
\begin{algorithm}[h]
	\linespread{1}\selectfont
	\caption{Diffusion management and tracking}\label{alg:bfsalg}
	\begin{algorithmic}[1]
		\For{$i=1,...,Nb_a$} \Comment {$Nb_a$: number of antennas}
		\For{$j=1,...,N_0$} \Comment {$N_0$: number of antenna trajectories}
		\State $Q_p$.enqueue($i$) \Comment{$Q_p$: queue for parent nodes}
		\State $V_d$.push\_back($j$) \Comment{$V_d$: vector of diffusion directions}
		\While{$k \leq M_d$} \Comment{$M_d$: maximum number of multiple diffusion}
		\For{$l=1,...,Q_p$}
		\For{$m=1,...,V_d$}
		\State locate $ImpactPoint$
		\State evaluate propagation information
		\State $Q_c$.enqueue($ImpactPoint$) \Comment{$Q_c$: queue for child nodes}
		\EndFor
		\EndFor
		\State $Q_p$.replace($Q_c$) \Comment{current child nodes become the new parent nodes}
		\State $Q_c$.clear() \Comment{clear queue and reclaim memory from old child nodes}
		\State $k=k+1$ \Comment{increment level of multiple diffusion}		
		\EndWhile
		\State $Q_p$.clear() \Comment{clear queue and reclaim memory from parent nodes}
		\EndFor
		\EndFor
	\end{algorithmic}
\end{algorithm}

\subsection{BRDF computation and storage}
\label{subsec:brdfcomp}
The main implementation challenges of the diffusion model are to efficiently compute, store and retrieve $\rho$ values, since they are needed at each impact point.

\textbf{BRDF implementation:} the mathematical functions required by He's model to compute the BRDF are not included in any internal C++ library (Lambert W function, Digamma, etc.). Various external open-source libraries were investigated like PARI, SageMath \cite{sagemath}, Boost \cite{boost}, etc. We selected PARI since it is written in C, contains all the required functions, and supports multi-precision computations.

\textbf{Memory management and optimization:} PARI initializes a stack in the memory where BRDF values are computed. The objective behind our implementation is to minimize memory usage and allow numerical simulations regardless of the system's performance. However, without any optimization, computing $\rho$ for a single incident angle consumes a huge amount of memory and eventually crashes when simulating for all $\theta_i$. This major issue was tackled by implementing a customized algorithm for manual memory management and optimization (Algorithm~\ref{alg:parigp}).
\begin{algorithm}[h]
	\linespread{1}\selectfont
	\caption{Optimized BRDF computation} \label{alg:parigp}
	\begin{algorithmic}[1]
		\State avma \Comment {default stack pointer}
		\State av \Comment {user-defined stack pointer}
		\State pari\_init(size) \Comment {open stack of size Bytes}
		\State av=avma \Comment {avma is initially at top of stack}
		\ForAll{($\theta_i$, $\theta_d$ and $\phi_d$)}
		\State compute $\rho(\theta_i,\theta_d,\phi_d,\sigma_0,\tau,\lambda,n_r,n_i)$
		\State avma=av \Comment {go back to top and overwrite}
		\EndFor
		\State stack.close() \Comment {close the PARI stack}
	\end{algorithmic}
\end{algorithm}

Objects are created in the PARI stack as $\rho$ is computed for each discretized direction of the diffusion plane. Since these directions are independent of each other, initialized objects at each iteration become useless once the corresponding $\rho$ value is computed. At each iteration, the GC algorithm reclaims the memory occupied by these `garbage' objects and overwrites each $\rho$ computation on the same memory resources, hence, minimizing PARI stack's requirements without impacting performance. Implementing this algorithm reduces memory consumption by a factor of 30: only a 100 MB stack (instead of 3 GB) is required to compute $\rho$ for all $\theta_i$, $\theta_d$ and $\phi_d$. $\rho$ is computed based on angular discretization steps $\Delta\theta_d$ and $\Delta\phi_d$, such that the discrete integration of $\rho$ over the diffusion space satisfies energy conservation, i.e.:
\begin{equation}
\sum_{\theta_d}\sum_{\phi_d}\rho\sin\theta_d\,\Delta\theta_d\Delta\phi_d\leq1
\label{eq:delta}
\end{equation}
Starting from $\Delta\theta_d=\Delta\phi_d=1^{\circ}$, their optimal values are automatically selected by decreasing them until the variation of (Eq.~\ref{eq:delta}) is less than an error threshold. Computing $\rho$ for a single incident angle, material and wavelength, takes 21 seconds for a $1^{\circ}$ discretization and up to 30 minutes for a $1/32^{\circ}$ discretization.

\textbf{Storing and retrieving BRDF values:} numerical simulations of $\rho$ are stored in databases using SQLite, a fast, full-featured and open-source C-written SQL database engine. Each database contains $\rho$ values for a given material, wavelength and for all $\theta_i$, $\theta_d$ and $\phi_d$. Inserting records in a database is generally done by executing an insert statement using the classical sql\_exec() routine since it requires minimal coding lines. While this method is adequate for inserting a single record, it is extremely inefficient when recursively inserting data in a database. This is due to the following facts: Firstly, sql\_exec() is a convenience wrapper around three functions: (i) sql\_prepare() compiles an SQL statement, (ii) sql\_step() executes a compiled statement and (iii) sql\_finalize() deletes a prepared statement.
Therefore, when using it recursively, the insert statement is compiled, executed and deleted at each iteration. Secondly, sql\_exec() passes SQL statements in transactions when executing them. Therefore, when using it recursively a new transaction gets opened and closed at each iteration. All these factors tremendously slow down a recursive insert operation, making the use of sql\_exec() highly inefficient: up to 8.5 hours for a $1^{\circ}$ discretization steps, for all $\theta_i$, $\theta_d$ and $\phi_d$ (equivalent to 2916000 $\rho$ values) with only 95 inserts/second, on the same low-end processor as in Section~\ref{subsec:indenv}.

An optimized insert operation was implemented to accelerate $\rho$ values storage (Algorithm~\ref{alg:sqlite}). Since the insert statement is the same for each iteration, it is compiled only once. This statement is then recursively executed by binding it to the corresponding $\theta_i$, $\theta_d$, $\phi_d$ and $\rho$ values. The execution process is done in a single transaction. The statement is then deleted as a final step.
\begin{algorithm}[h]
	\linespread{1}\selectfont
	\caption{Optimized recursive insert procedure} \label{alg:sqlite}
	\begin{algorithmic}[1]
		\State stmt = insert values in DB \Comment {insert statement}
		\State prepare(stmt) \Comment {compile statement}
		\State exec(transaction.begin) \Comment {open transaction}
		\ForAll{($\theta_i$, $\theta_d$ and $\phi_d$)}
		\State compute $\rho(\theta_i,\theta_d,\phi_d,\sigma_0,\tau,\lambda,n_r,n_i)$
		\State stmt = sql.bind(values) \Comment {bind values}
		\State step(stmt) \Comment {execute stmt}
		\State bindings.clear(stmt) \Comment {clear value bindings}
		\State reset(stmt) \Comment {reset execution flag}
		\EndFor
		\State exec(transaction.end) \Comment {close transaction}
		\State exec(index.create) \Comment {indexing table}
		\State finalize(stmt) \Comment {delete stmt}
	\end{algorithmic}
\end{algorithm}

Although this method is much more complex than the classical one, it reduces the insert time by a factor of 9: storing the same amount of $\rho$ values using the same processor takes approximately 1 hour with 860 inserts/second, and generates a database of size 245 MB. As for retrieving values from the database, a huge improvement was observed using this optimized method, where a select statement for a single value of $\rho$ takes no longer than 2 ms, compared to approximately 1000 ms with sql\_exec().

\textbf{Deducing the uniform diffusion component:} the difference $1-(\rho_{sr}+\rho_{dd}+\rho_{ud})$ is the fraction of transmitted energy $\rho_{tr}$. He's model does not provide an expression for $\rho_{ud}$; the two unknown components ($\rho_{ud}$ and $\rho_{tr}$) are deduced by performing a trial and error analysis of the Fresnel transmission coefficient behavior when independently varying $\theta_i$, $n_i$ and $\sigma_0$. This method provides approximate values for each component which cannot be accurately verified, but respect energy conservation.  

\subsection{Managing diffusion directions}
\label{subsec:diffmanage}
Another major implementation challenge is managing the potential explosion of diffusion directions: with multiple diffusion, the number of trajectories to process becomes huge. Various techniques can be used to tackle this challenge and reduce simulation time and memory usage.

First, a relative threshold can be imposed on $\rho$ as a function of $\rho_{max}(\theta_i)$. This eliminates all the directions with a $\rho$ value inferior to this threshold; hence, reducing both simulation time and memory usage. Thresholds can also be imposed on power values, prohibiting further diffusion at an impact point. However, power thresholds should be set such that they don't eliminate major propagation directions from the first diffusion. Moreover, using generated $\rho$ databases, discretization steps $\Delta\theta_d$ and $\Delta\phi_d$ can be increased while still verifying energy conservation (Eq.~\ref{eq:delta}). This reduces the number of diffusion directions and hence, the number of iterated $\rho$ values, without the need to generate another database.

\section{First Simulations Performance Results}
\label{sec:results}
\begin{figure}[h]
	\centering
	\includegraphics[width=11.5 cm]{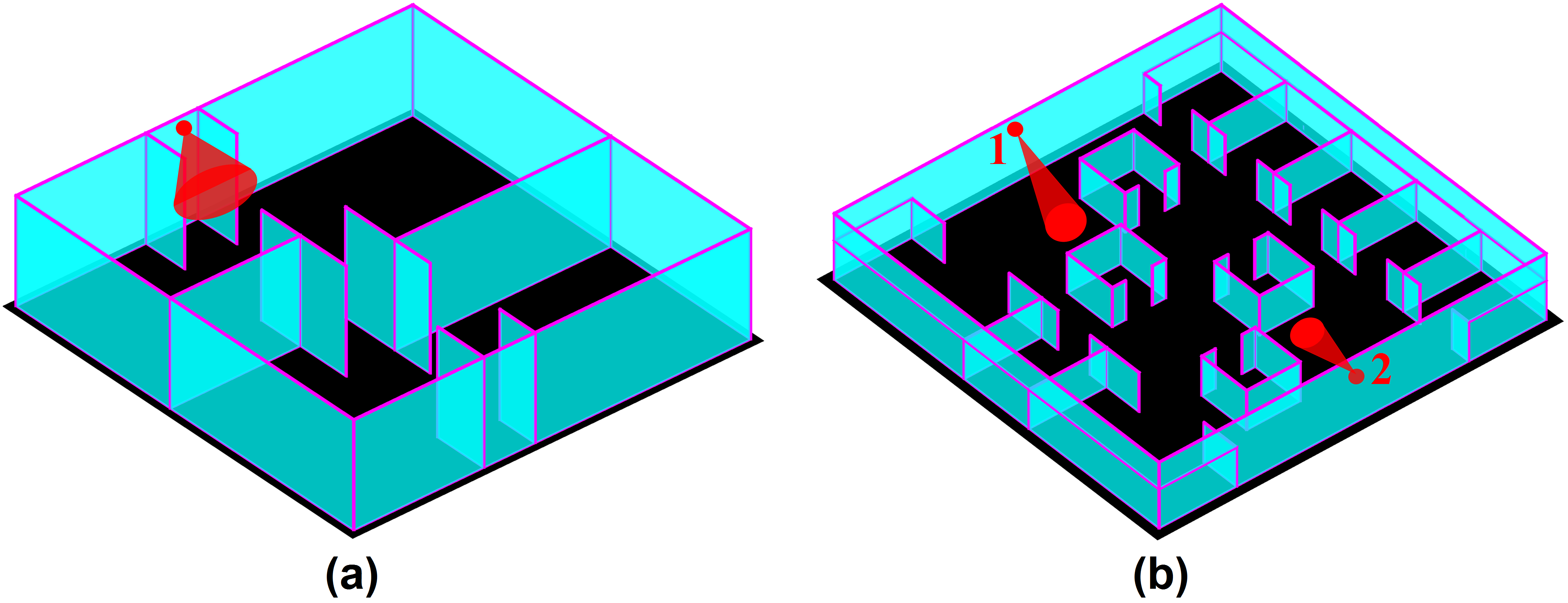}
	\caption{A rotated 3D view of the two simulation environments. (a) FWA - typical apartment with a large aperture antenna, (b) i4.0 - typical factory with 2 narrow aperture antennas.}
	\label{fig:simenvir}
\end{figure}

We present in this section the simulation results for two major 5G indoor applications: Fixed Wireless Access (FWA) in an apartment at 60 GHz and industry 4.0 (i4.0) in a factory at 26 GHz (Fig.~\ref{fig:simenvir}). Four scenarios are considered (Table~\ref{tab:scenarios}) that allow us to compare the contribution of reflection, single and double diffusion on the indoor propagation simulation of 5G mmWaves, where diffusion is the dominant mechanism. These scenarios intend to test the performance of iGeoStat when simulating multiple diffusion in realistic environments and its ability to generate various output maps. Simulation results are then validated by comparing them to existing measurement campaigns at 60 GHz and 26 GHz.

\begin{table}[h]
	\caption{Simulations scenarios.}
	\label{tab:scenarios}
	\centering
	\begin{tabular}{|c|c|c|}
		\hline
		\textbf{Use Case} & \textbf{Studied Mechanisms} & \textbf{Antenna Configuration}\\
		\hline
		FWA & reflection vs. single diffusion & 1 antenna with a large angular aperture ($\pi$ sr)\\
		\hline
		i4.0 & single vs. double diffusion & 2 antennas with narrow angular apertures ($\frac{\pi}{1000}$ sr)\\
		\hline 
	\end{tabular} 
\end{table}
The FWA scenarios are run on a mid-range hardware configuration: a 2.4 GHz Intel Core i5-6300U CPU with 8 GB of RAM, while the i4.0 ones are run on a more advanced configuration: a 2.2 GHz Intel Xeon E5-2699v4 CPU with 532 GB of RAM.

Using measurements from \cite{Sato} and \cite{Li}, three BRDFs (Table~\ref{tab:BRDF}) were generated for the following angles: $\theta_i\in[0,\frac{\pi}{2}]$, $\theta_d\in[0,\frac{\pi}{2}]$ and $\phi_d\in[0,2\pi]$ with $\Delta\theta_d=\Delta\phi_d=1^{\circ}$. This corresponds to 2916000 $\rho$ values per BRDF, computed and stored in database tables using the advanced hardware configuration in 40 minutes with 1215 inserts/second, and is done only once before launching the simulations in iGeoStat. The surface roughness parameters $\sigma_0$ and $\tau$ are chosen such that the diffusion profile corresponds to a moderately rough surface (Fig.~\ref{fig:visubrdf}(b)).

To test the full potential of our implementation, no thresholds are imposed on $\rho$ nor on power, and all the propagation information are computed and stored for each impact.
\begin{table}[h]
	\caption{Computed BRDFs.}
	\label{tab:BRDF}
	\centering
	\begin{tabular}{|c|c|c|c|c|c|c|}
		\hline
		\textbf{BRDF} & \textbf{Material} & \textbf{f} (GHz) & \bm{$\lambda$} (cm) & \bm{$\sigma_0$} (cm) & \bm{$\tau$} (cm) & \bm{$n_c$}\\
		\hline
		$\rho_{p,60}$ & Plasterboard (A) & 60 & 0.52 & 0.3 & 2.8 & $1.76-0.016i$\\
		\hline
		$\rho_{p,26}$ & Plasterboard & 26 & 1.13 & 0.6 & 3 & $1.82-0.117i$\\
		\hline
		$\rho_{c,26}$ & Concrete & 26 & 1.13 & 0.6 & 3 & $1.21-0.256i$\\
		\hline 
	\end{tabular} 
\end{table}

\subsection{FWA - reflection vs. single diffusion}
In this use case, we first impose a reflection-only simulation to study whether this mechanism can illustrate or not the actual propagation behavior of 5G mmWaves in an indoor environment. We then introduce single diffusion and compare its contribution on simulating indoor mmWave propagation to the reflection-only scenario.

\textbf{Indoor environment and materials:} using the custom mode in iGeoStat, we generate an apartment with a total area of 150 m$^2$ where all the walls go up to the ceiling of height $h_c=3$~m (Fig.~\ref{fig:simenvir}(a)). All the obstacles (walls, floor, and ceiling) are covered in ``Plasterboard (A)''~\cite{Sato}, configured with a moderately rough surface profile ($\sigma_0/\lambda=0.57$).

Imposing a reflection-only scenario is equivalent to considering that all obstacles have a smooth surface ($\sigma_0=0$), which does not reflect the physical properties of our environment. In this case, the BRDF corresponds to computing the Fresnel coefficient for specular reflection directions only, where $\theta_d=\theta_i=[0,\frac{\pi}{2}]$ and $\phi_d=0$ with $\Delta\theta_d=1^{\circ}$. This corresponds to 90 $\rho$ values, computed and stored in a database table in 0.1 seconds. For the single diffusion scenario, the actual physical environment properties are taken into account through BRDF $\rho_{p,60}$ (Table~\ref{tab:BRDF}).

\textbf{Antenna and measurement plane:} An antenna is placed on the main corridor's left wall at $h_a=2.8$ m (Fig.~\ref{fig:simenvir}(a)). It emits $P_0=~20$~mW in a large angular aperture $\Omega_a=\pi$ sr, where   $\theta_0\in[-90^{\circ},90^{\circ}]$ and $\phi_0\in[90^{\circ},180^{\circ}]$ are centered in the direction ($\theta_0^a=0^{\circ},\phi_0^a=135^{\circ}$), pointing to the floor in the $\vec{x}$ direction. $\Omega_a$ is discretized according to $\Delta\theta_0=\Delta\phi_0=1^{\circ}$, which corresponds to $N_0=16200$ trajectories ($\Omega_a=16200\;d^2\Omega_0$). The measurement plane is set at $h_m=1.2$ m above the floor and is divided for precision purposes into a relatively high $100\times100$ grid resolution with respect to our environment. 

For the reflection scenario, each of the $N_0$ trajectories is set to undergo a very high number (1000) of reflections, i.e. interceptions by indoor obstacles. As for the single diffusion scenario, each of the $N_0$ trajectories hits one indoor obstacle and then undergoes diffusion.

\begin{figure}[h]
	\centering
	\includegraphics[width=13.6 cm]{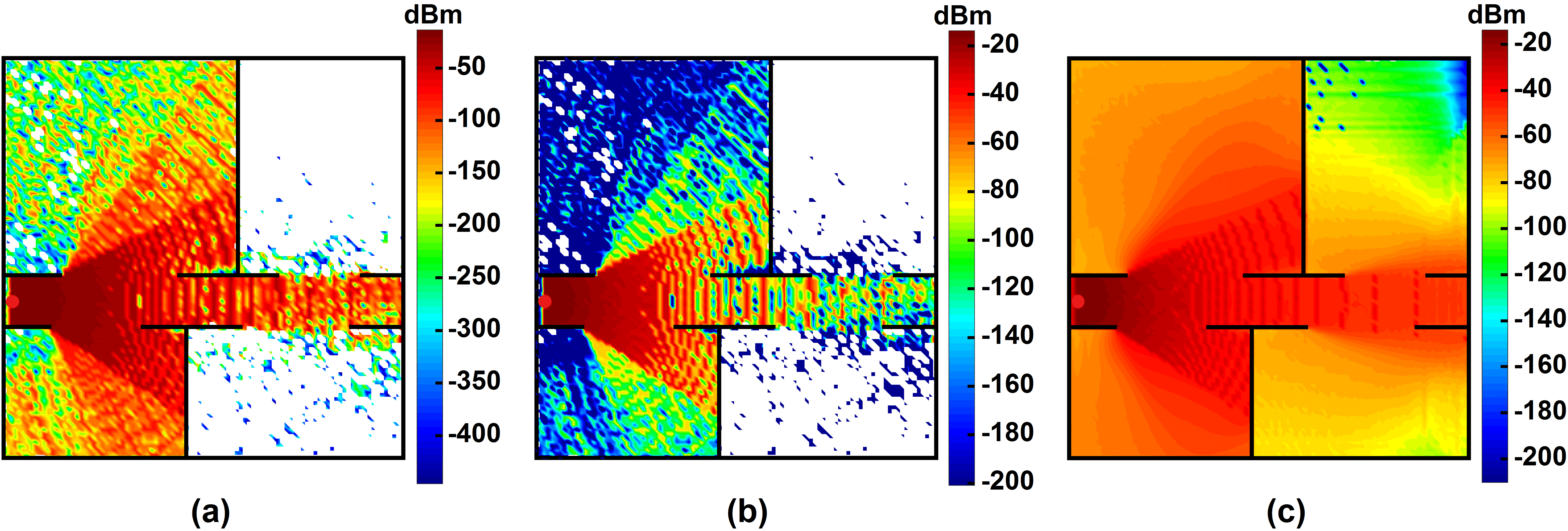}
	\caption{FWA power maps. (a) Reflection scenario showing large coverage holes. (b) Filtered reflection map at -200 dBm. (c) Single diffusion scenario showing full environment coverage.}
	\label{fig:FWAres1}
\end{figure}

\textbf{Simulation output:} the power map of the reflection scenario (Fig.~\ref{fig:FWAres1}(a)) shows multiple coverage holes in the environment. For convenience, received power values are expressed in dBm, but do not have any physical meaning since such low values cannot be measured. Filtering at -200 dBm (Fig.~\ref{fig:FWAres1}(b)) shows a 20\% environment coverage only. On the other hand, using the same antenna configuration, the power map of single diffusion (Fig.~\ref{fig:FWAres1}(c)) shows a fully covered environment, with typical received power varying between -208 and -17 dBm.

These power maps highlight the crucial contribution of diffusion on indoor mmWave propagation and its advantage over reflection when simulating the interaction between EM waves and materials, especially if properly modeled and implemented. This shows that a classical reflection-only scenario cannot reflect the actual indoor behavior of 5G mmWaves even with a huge number of reflections. Hence, existing classical ray tracing techniques that mainly employ reflection are not adapted to study indoor 5G mmWave propagation.

\begin{figure}[h]
	\centering
	\includegraphics[width=9.2 cm]{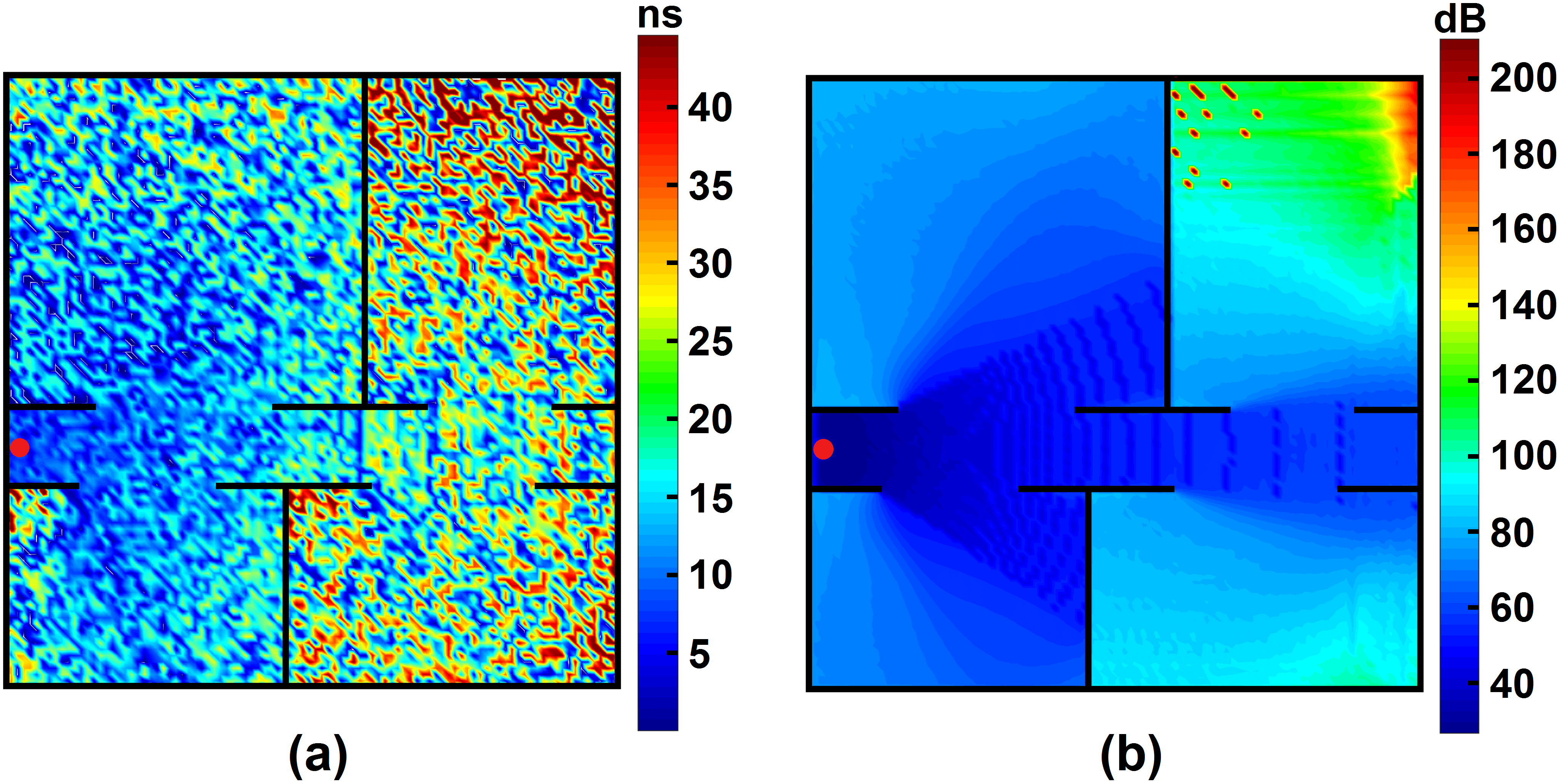}
	\caption{FWA additional maps for single diffusion. (a) Delay spread. (b) Path loss variability.}
	\label{fig:FWAres2}
\end{figure}

Additional output maps can be generated in iGeoStat for the single diffusion scenario, based on the propagation information stored by the impacts. Fig.~\ref{fig:FWAres2}(a) is a delay spread map that shows the average arrival time difference of the various multipath trajectories in each grid element. Values vary between 0.5 ns close to the antenna and 44 ns in its far region. Fig.~\ref{fig:FWAres2}(b) is a path loss variability map that allows to visualize how power decreases in the environment, showing an analogy to Fig.~\ref{fig:FWAres1}(c). Values vary between 29 and 215 dB as we move away from the antenna.

\subsection{i4.0 - single vs double diffusion}
Previous scenarios showed that diffusion must be taken into account when simulating indoor mmWave propagation. However, some cases might exist where single diffusion is insufficient to predict mmWaves behavior. Therefore, we first test this mechanism in a different environment geometry and antenna configuration. We then introduce double diffusion and compare its contribution on simulating indoor mmWave propagation to the single diffusion scenario.

\textbf{Indoor environment and materials:} using the custom mode in iGeoStat, we generate a factory with a total area of 1000 m$^2$ (Fig.~\ref{fig:simenvir}(b)). The ceiling is at $h_c=5$ m, however the walls' height is fixed at $h_{wa}=3$ m, allowing radio waves to propagate in the whole environment.

The walls are covered in ``Plasterboard'' \cite{Li} with BRDF $\rho_{p,26}$. The floor and the ceiling are covered in ``Concrete'' \cite{Li} with BRDF $\rho_{c,26}$. Both materials are configured with a moderately rough surface profile ($\sigma_0/\lambda=0.53$).

\textbf{Antenna and measurement plane:} two antennas are placed on opposite walls of the environment at $h_a=4.8$ m (Fig.~\ref{fig:simenvir}(b)). Both antennas emit $P_0=200$ mW in a narrow angular aperture $\Omega_a=\frac{\pi}{1000}$ sr divided into $N_0=16$ trajectories ($\Omega_a=16\;d^2\Omega_0$) with $\Delta\theta_0=\Delta\phi_0=1^{\circ}$. For antenna 1, $\theta_0\in[-2^{\circ},2^{\circ}]$ and $\phi_0\in[133^{\circ},137^{\circ}]$ are centered in the direction ($\theta_0^a=0,\phi_0^a=135^{\circ}$), pointing to the floor in the $\vec{x}$ direction. For antenna 2, $\theta_0\in[178^{\circ},-178^{\circ}]$ and $\phi_0\in[133^{\circ},137^{\circ}]$ are centered in the direction ($\theta_0^a=180^{\circ},\phi_0^a=135^{\circ}$), pointing to the floor in the $-\vec{x}$ direction. The measurement plane is set at $h_m=1.8$ m and is divided into an adapted $100\times100$ grid resolution with respect to our environment.

For the single diffusion scenario, every trajectory of each antenna undergoes diffusion after hitting one obstacle only. For the double diffusion scenario, each of the resulting diffusion directions is allowed to hit and diffuse on an additional obstacle.

\begin{figure}[h]
	\centering
	\includegraphics[width=16.5 cm]{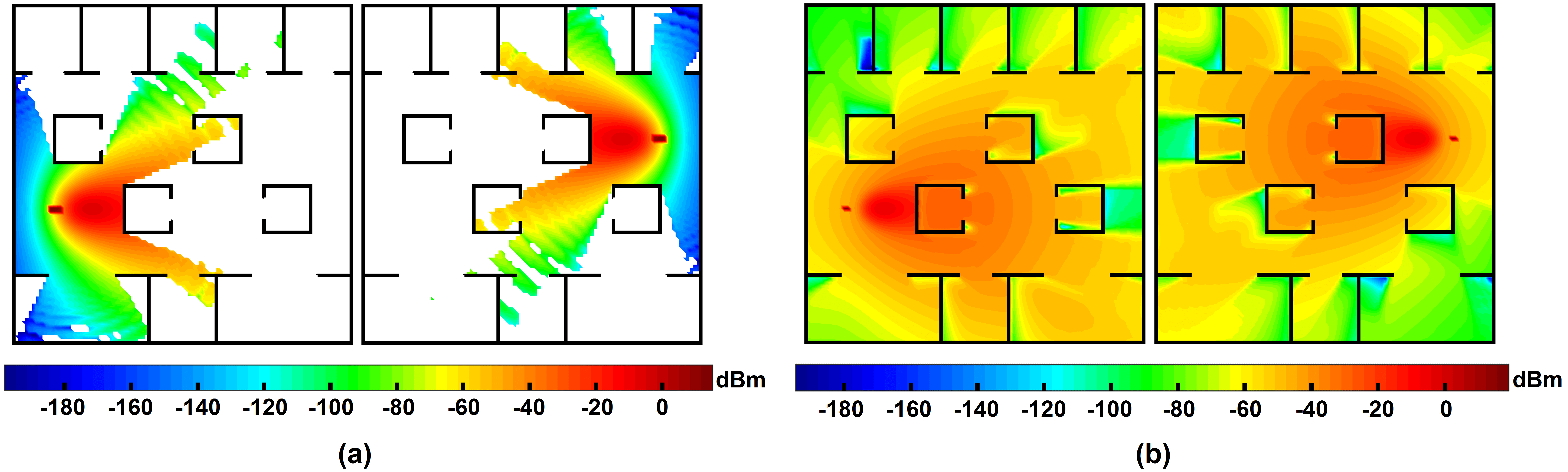}
	\caption{i4.0 power maps. (a) Single diffusion scenario showing large coverage holes for both antennas. (b) Double diffusion scenario showing a fully covered environment for both antennas.}
	\label{fig:I40res1}
\end{figure}

\textbf{Simulation outputs:} the output power maps for the single diffusion scenario (Fig.~\ref{fig:I40res1}(a)) show typical values varying between -197 and 12 dBm for each antenna. However, large coverage holes are shown, with an only 35\% environment coverage by each antenna separately. When placing both antennas together, the resulting coverage corresponds to 58\% of the whole environment. Introducing double diffusion has a major effect on coverage as shown in the output maps of both antennas (Fig.~\ref{fig:I40res1}(b)). They illustrate a fully covered environment by each antenna separately with a very slight increase in received power that vary between -190 and 18 dBm.\\

These results confirm that in some cases, simulating single diffusion is not sufficient to study indoor mmWave propagation; multiple diffusion simulation would be required to properly illustrate the indoor behavior of 5G mmWaves. These results also show that the main power is concentrated in the first diffusion level, which gives an idea regarding the adequate level of multiple diffusion simulation for a given scenario.

\begin{figure}[h]
	\centering
	\includegraphics[width=9 cm]{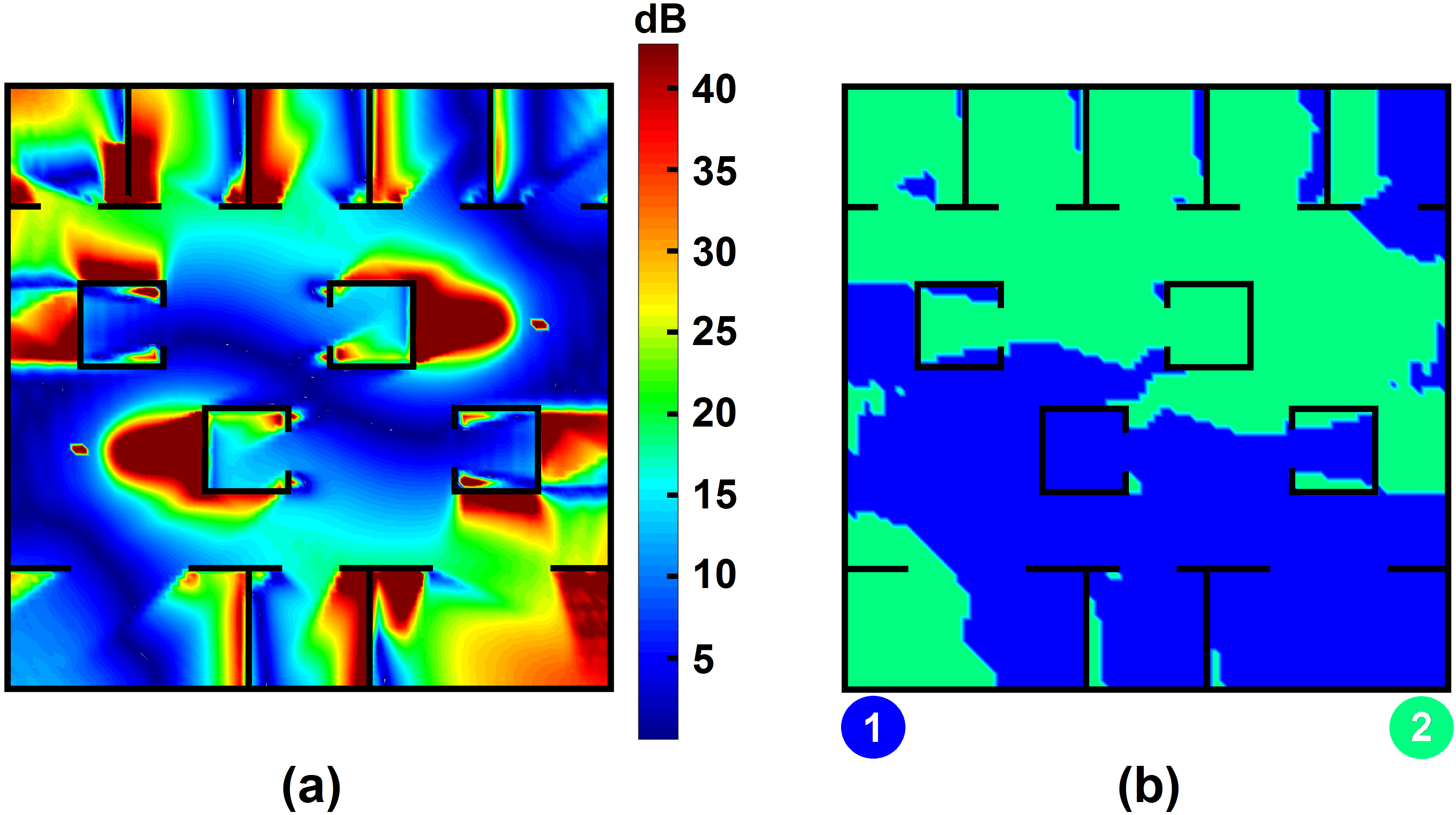}
	\caption{i4.0 additional maps for double diffusion scenario. (a) Resulting SINR map, having both antennas in the environment. (b) Optimal coverage areas for antennas 1 (blue) and 2 (green).}
	\label{fig:I40res2}
\end{figure}

Having multiple antennas in the environment, additional output maps can be generated in iGeoStat for the double diffusion scenario. Fig.~\ref{fig:I40res2}(a) is the resulting SINR map when placing both antenna in the environment. Values vary between 0.2 and 48 dB. The coverage map in Fig.~\ref{fig:I40res2}(b) illustrates the optimal coverage areas by each antenna, and shows that antenna 1 (blue) optimally covers 54.6\% of the factory compared to 45.4\% for antenna 2 (green).

\subsection{Results validation}
As a first results validation, we generate the LOS and NLOS path loss $PL(dB)$ curves with respect to the transmitter-receiver distance $r(m)$ for both use cases and compare them to the ones obtained from existing measurements at 60 GHz \cite{Kacou} and 26 GHz \cite{Rubio, Pimienta}. We note that these measurements are conducted in different environments and antenna configurations from ours. Hence, we are more interested in comparing the trend of the path loss curves rather than their exact values, as more simulations are needed to characterize path loss for a given scenario. 

\begin{figure}[h]
	\centering
	\includegraphics[width=16.5 cm]{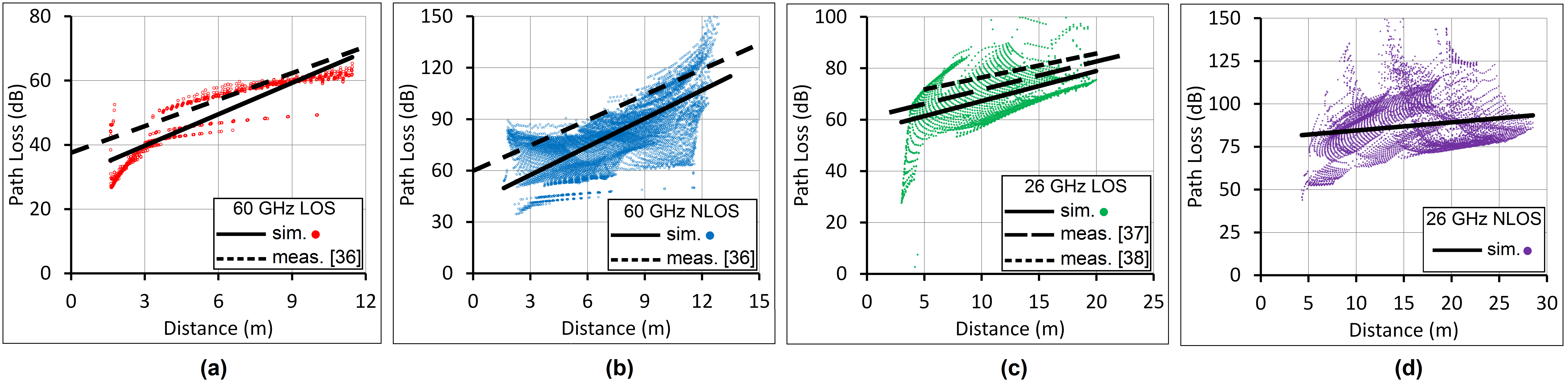}
	\caption{Path loss curves based on results from our simulations (sim.) and from measurements (meas.). (a) 60 GHz LOS, (b) 60 GHz NLOS, (c) 26 GHz LOS, (d) 26 GHz NLOS.}
	\label{fig:pathloss}
\end{figure}

\textbf{60 GHz use case:} the power map in Fig.~\ref{fig:FWAres1}(c) are used to generate the path loss curves by fitting 10000 positions based on linear regression, without any filtering. For the LOS scenario (Fig.~\ref{fig:pathloss}(a)), the plotted curve $PL=3.2\;r+30$ is based on 1648 positions (red dots). The NLOS scenario (Fig.~\ref{fig:pathloss}(b)) has a higher and steeper path loss curve $PL=5.5\;r+41$ and is based on 8352 positions (blue dots). The resulting path loss curves from measurements in \cite{Kacou} are based on fitting 21 combined positions for LOS and NLOS.

\textbf{26 GHz use case:} the power maps in Fig.~\ref{fig:I40res1}(b) are used to generate the path loss curves by fitting 20000 positions (10000 for each antenna) based on linear regression, without any filtering. For the LOS scenario (Fig.~\ref{fig:pathloss}(c)), the plotted curve $PL=1.2\;r+55$ is based on 6272 positions (green dots). The NLOS scenario (Fig.~\ref{fig:pathloss}(d)) has a higher path loss curve $PL=0.5\;r+80$ with a more gentle slope and is based on 13728 positions (purple dots). The resulting path loss curves from measurement in \cite{Rubio} and \cite{Pimienta} are based on 10 and 264 positions respectively for LOS only as they do not consider NLOS.

\textbf{Discussion:} Very positive results are shown by the generated path loss curves from our simulations (sim.), showing very similar trendlines to the ones obtained from measurements (meas.). The slight gap between the obtained values is mainly due to the difference in environments geometry and antenna configurations, especially the transmitting and receiving antenna gains which are considered in all the (meas.) path loss curves. The difference in values even between the two LOS (meas.) curves of \cite{Rubio} and \cite{Pimienta} (Fig.~\ref{fig:pathloss}(c)) confirms that diffusion is strongly impacted by the environment settings which leads to various indoor mmWave propagation behavior. Hence, one measurement campaign is also insufficient to characterize path loss for a given scenario.

\subsection{Performance results}
These simulation results are very promising; they highlight the ability of our system implementation to simulate multiple diffusion in a reasonable amount of resources and to generate various output maps that illustrate the actual indoor behavior of 5G mmWave propagation.

We note that when simulating any propagation mechanism (reflection, single, double diffusion, etc.), the total RAM usage is equal to the amount used by a single antenna trajectory, independently of the number of antennas and trajectories $N_0$. This is due to the efficiency of Algorithm~\ref{alg:bfsalg} where the same memory resources are recycled and used over and over again. This allows us to anticipate the amount of required RAM prior to the simulation.    

The performance results are presented in Table~\ref{tab:performance}. Although FWA scenarios are simulated on a mid-range hardware configuration with $N_0=16200$ trajectories, simulation times are very reasonable and RAM usage is extremely low considering the huge amount of registered impacts, containing all the propagation information. i4.0 scenarios were simulated on a more advanced hardware configuration with $N_0=16$ trajectories; for the single diffusion case, simulation time is very fast and memory usage is extremely low and is as expected, equal to the FWA scenario, despite a huge difference in number of impacts. As stated above, this is due to the efficiency of Algorithm~\ref{alg:bfsalg}. For the double diffusion scenario, simulation time and RAM usage are both relatively high compared to the previous scenarios. As expected, this mechanism is extremely complex due to the explosion of diffusion directions and consequently the number of impacts.
\begin{table}[h]
	\caption{Performance indicators per antenna.}
	\label{tab:performance}
	\centering
	\begin{tabular}{|c|c|c|c|c|}
		\hline
		\textbf{Performance Indicator} & \textbf{FWA - Reflection} & \textbf{FWA - 1 Diffusion} & \textbf{i4.0 - 1 Diffusion} & \textbf{i4.0 - 2 Diffusion}\\
		\hline 
		Number of impacts & $16.2\times10^6$ & $\approx 524.9\times10^6$ & $\approx 518.4\times10^3$ & $\approx 17.6\times10^9$\\
		\hline
		Simulation time & 1.5 hours & 2.5 hours & 20 seconds & 60 hours\\
		\hline 
		RAM usage & 2 MB & 10 MB & $\approx$ 10 MB & $\approx$ 230 GB\\ 
		\hline 
	\end{tabular} 
\end{table}

Although simulation time and RAM usage for double diffusion might seem high, it would have been impossible to simulate this mechanism without the implemented data structure and the Garbage Collection integrated in Algorithm~\ref{alg:bfsalg}. Both RAM usage and simulation time can be reduced by storing only the required propagation information for a simulation's objectives; thus, minimizing used memory per impact. Moreover, optimization techniques presented in Section~\ref{subsec:diffmanage}, like filtering diffusion directions whose power is below thermal noise, reduces the number of impacts. These techniques will also allow us to go beyond double diffusion if needed in a reasonable amount of resources.

Generating such output maps as in iGeoStat would have been impossible with classical ray tracing techniques that mainly simulate reflection. Including diffusion would create many complicated challenges like power evaluation from multiple rays and selecting/verifying the appropriate number of rays that ensures the representation of the whole diffusion lobe. This would correspond to computing the BRDF at each impact, which requires huge computational resources. In our system implementation, the BRDF is computed only once prior to a simulation, and is stored in a database which can be used for other simulations.

\section{Conclusion and Future Work}
\label{sec:conclusion}
This paper introduced a novel modeling framework for indoor 5G mmWave propagation, combining stochastic environment modeling with physical propagation simulation. Its system implementation, iGeoStat, generates parameterized typical environments that account for the statistical indoor variability, then simulates radio propagation based on the physical interaction between EM waves and indoor materials. Challenging computational tasks were solved by formulating an adapted link budget and designing new memory management and optimization algorithms, making iGeoStat the first to simulate multiple diffusion in realistic environments.

First simulations were carried for FWA at 60 GHz and i4.0 at 26 GHz to investigate the contribution of reflection, single and double diffusion on indoor mmWave propagation simulation, where diffusion is the most complex but certainly the most essential mechanism. Illustrated results confirmed that a reflection-only simulation cannot describe indoor mmWaves behavior. They also showed that received power is mostly concentrated in single diffusion; however, simulating this mechanism is not always sufficient to study indoor mmWave coverage, and may require double diffusion. Such conclusions cannot be obtained with classical ray tracing techniques.

Simulation results were successfully validated by comparing the generated LOS and NLOS path loss curves to the ones obtained from measurements at 60 and 26 GHz. The performance results showed the efficiency of iGeoStat to simulate multiple diffusion and generate various output maps (power, SINR, coverage, path loss and delay spread) in a reasonable amount of time and memory. Our implementation enables us to determine the impact of a given multiple diffusion level and to predict the required time and RAM for any propagation mechanism.

Optimization techniques presented in this paper can further reduce simulation time and memory usage, allowing us to go beyond double diffusion if needed. iGeoStat's flexibility will enable us to implement additional modules (indoor furniture, antenna placement optimization, throughput analysis, etc.) to solve indoor planning and dimensioning 5G mmWaves challenges. Extensive statistical studies on simulations results will be carried for various sets of parameterized stochastic environments (geometry and materials) to propose the first analytical formulae of indoor propagation models that are not dedicated to a particular environment, frequency or use case.

\end{document}